\documentclass[aps,prc,twocolumn,superscriptaddress]{revtex4-2}

\usepackage[T1]{fontenc}

\usepackage{graphicx}
\usepackage{xcolor}
\newcommand{\tssc}[1]{\textsuperscript{#1}}

\DeclareUnicodeCharacter{3B1}{\ensuremath{\alpha}}
\DeclareUnicodeCharacter{2212}{-}

\begin{document}


\title{Measurement of the \tssc{14}N(n,p)\tssc{14}C cross section at the CERN n\_TOF facility from sub-thermal energy to 800 keV}





\author{Pablo~Torres-S\'{a}nchez} \email{pablotorres@ugr.es}  \affiliation{University of Granada, Spain}%
\author{Javier~Praena} \affiliation{University of Granada, Spain} %
\author{Ignacio~Porras} \affiliation{University of Granada, Spain} %
\author{Marta~Sabat\'{e}-Gilarte} \affiliation{European Organization for Nuclear Research (CERN), Switzerland} \affiliation{Universidad de Sevilla, Spain} %
\author{Claudia~Lederer-Woods} \affiliation{School of Physics and Astronomy, University of Edinburgh, United Kingdom} %
\author{Oliver~Aberle} \affiliation{European Organization for Nuclear Research (CERN), Switzerland} %
\author{Victor~Alcayne} \affiliation{Centro de Investigaciones Energ\'{e}ticas Medioambientales y Tecnol\'{o}gicas (CIEMAT), Spain} %
\author{Simone~Amaducci} \affiliation{INFN Laboratori Nazionali del Sud, Catania, Italy} \affiliation{Dipartimento di Fisica e Astronomia, Universit\`{a} di Catania, Italy} %
\author{J\'{o}zef~Andrzejewski} \affiliation{University of Lodz, Poland} %
\author{Laurent~Audouin} \affiliation{Institut de Physique Nucl\'{e}aire, CNRS-IN2P3, Univ. Paris-Sud, Universit\'{e} Paris-Saclay, F-91406 Orsay Cedex, France} %
% ...
%

\collaboration{The n\_TOF Collaboration (www.cern.ch/ntof)} \noaffiliation

\date{\today}

\begin{abstract}

\begin{description}
\item[Background] 
The \tssc{14}N(n,p)\tssc{14}C reaction is of interest in neutron capture therapy, where nitrogen-related dose is the main component due to low-energy neutrons, and in astrophysics, where \tssc{14}N acts as a neutron poison in the s-process. Several discrepancies remain between the existing data obtained in partial energy ranges: thermal energy, keV region and resonance region.
\item[Purpose] 
Measuring the \tssc{14}N(n,p)\tssc{14}C cross section from thermal to the resonance region in a single measurement for the first time, including characterization of the first resonances, and providing calculations of Maxwellian averaged cross sections (MACS).
\item[Method] 
Time-of-flight technique. Experimental Area 2 (EAR-2) of the neutron time-of-flight (n\_TOF) facility at CERN. \tssc{10}B(n,$\alpha$)\tssc{7}Li and \tssc{235}U(n,f) reactions as references. Two detection systems running simultaneously, one on-beam and another off-beam. Description of the resonances with the R-matrix code \textsc{sammy}. 
\item[Results]
 The cross section has been measured from sub-thermal energy to 800 keV resolving the two first resonances (at 492.7 and 644 keV). A thermal cross-section (1.809$\pm$0.045 b) lower than the two most recent measurements by slightly more than one standard deviation, but in line with the ENDF/B-VIII.0 and JEFF-3.3 evaluations has been obtained. A 1/v energy dependence of the cross section has been confirmed up to tens of keV neutron energy. The low energy tail of the first resonance at 492.7 keV is lower than suggested by evaluated values, while the overall resonance strength agrees with evaluations. 
\item[Conclusions] 
Our measurement has allowed to determine the \tssc{14}N(n,p) cross-section over a wide energy range for the first time. We have obtained cross-sections with high accuracy (2.5 \%) from sub-thermal energy to 800 keV and used these data to calculate the MACS for $kT$ = 5 to $kT$ = 100 keV. 
\end{description}

\end{abstract}


\maketitle

\section{INTRODUCTION}

The \tssc{14}N(n,p)\tssc{14}C reaction plays a key role in many fields. It is relevant in nuclear astrophysics because it is a significant neutron poison in the s-process nucleosynthesis \cite{Wallner16}. Also, this reaction is an important proton feeder in the production of fluorine, via the chain of reactions \tssc{18}O(p,$\alpha$)\tssc{15}N($\alpha$,$\gamma$)\tssc{19}F. \tssc{19}F is a useful tracer of the physical conditions in stellar interiors, since it can be easily destroyed by proton or $\alpha$ reactions, hence any production site needs also to enable \tssc{19}F to escape from the stellar interiors as it is observed with high abundance in AGB atmospheres \cite{Abia15}. 
In radiotherapy, the dose in healthy tissues is a limiting factor. Specifically, in Boron Neutron Capture Therapy (BNCT), the \tssc{14}N(n,p)\tssc{14}C reaction constitutes the main contribution to the dose in healthy tissue due to thermal neutrons \cite{Goorley02}, therefore, it is crucial for determining the delivered dose in a BNCT treatment \cite{Barth12}. In addition, the International Commission on Radiation Units and Measurements (ICRU) recommends that the delivered dose should have less than 5\% deviation from the prescribed dose \cite{ICRU46}. At present, nuclear data are under review and upgrade for the new IAEA Technical Document (TECDOC) on BNCT \cite{tecdoc2022}. An accurate \tssc{14}N(n,p)\tssc{14}C cross section is one of most important data to be considered there. 


The \tssc{14}N(n,p)\tssc{14}C reaction has been measured several times, though all of these measurements have been focused on a specific energy region: thermal energy (25.3 meV), 1/v region (up to a few tens of keV), integral measurements in the keV range (25-200 keV) or resonance region (above 400 keV). Thus, there is no single measurement connecting the thermal energy or the 1/v region with the resonance region. The first resonance is above 450 keV. Despite the number of experiments there are some issues that need to be clarified in the different energy ranges. 

At thermal energy, there are several measurements \cite{Batchelor49,Coon49,Cuer51,Hanna61,Gledenov93,Wagemans00,Kitahara19} with values from 1.7 to 2.0 b. The last two provided ($1.93\pm0.05$ b) \cite{Wagemans00} and ($1.868\pm0.006$ b) \cite{Kitahara19} and they are higher than the ENDF/B-VIII.0 and JEFF-3.3 evaluations (1.8271 b) \cite{ChadwickEval99}. In BNCT, the disparities between measurements and evaluations in involved reactions can lead to differences up to 15 \% in the thermal neutron dose estimation in human tissues and 3-4 \% in total dose in no-tumor tissues (including $\gamma$ and boron-induced doses). Thus, a reduction of the uncertainty in the \tssc{14}N(n,p) reaction leads to more accurate treatments approaching to ICRU recommendations \cite{ICRU46}. The thermal value is also relevant in astrophysics. For instance, it has been shown that the thermal value impacts considerably on the Maxwellian averaged cross section (MACS) for (n,p) reactions \cite{Druyts94}. 


The only differential measurement available in the 1/v region is by Koehler \textit{et al.} \cite{Koehler89} that covers the range from 61 meV to 35 keV. It was normalized by extrapolation to the thermal value in the Nuclear Data Compilation by Ajzenberg-Selove, at $1.83\pm0.03$ b \cite{Ajzenberg-Selove86}. In a second measurement, Koehler \textit{et al.} confirmed the assumed value at thermal energy \cite{Koehler93}, obtaining the same results as Gledenov \textit{et al.} \cite{Gledenov93}. Between 35 and 150 keV, there are no differential measurements, and there is only derived data from the inverse-reaction measurement by Gibbons \textit{et al.} \cite{Gibbons59}.


In the astrophysical range several integral measurements have been carried out  \cite{Brehm88,Gledenov94,Sanami97,Shima97,Wallner16}. For all of them there is a good agreement at 25 keV, except Brehm \textit{et al.} \cite{Brehm88}. 
However, some differences arise in the measurements above 25 keV. Shima \textit{et al.} \cite{Shima97} observed a reduction in the cross-sections from 1.67 mb at 35.8 keV to 1.19 mb at 67.1 keV, while Gledenov \textit{et al.} \cite{Gledenov94} saw a mostly flat behavior between 24.5 keV and 144 keV with values oscillating between 2.04 and 2.08 mb. The last measurement by Wallner \textit{et al.} \cite{Wallner16} encountered a reduction above 25 keV, with values of 0.88 and 0.90 mb at 123 and 178 keV, respectively, in clear disagreement with Gledenov \textit{et al.} \cite{Gledenov94}.

In the resonance region the most important measurements were carried out by Johnson \textit{et al.} \cite{Johnson50} and Morgan \cite{Morgan79}. Johnson \textit{et al.} measured the cross section from 150 keV to 2.15 MeV, using neutrons from a lithium target bombarded by protons with an energy spread of 5 keV \cite{Johnson50}. Later, Morgan \cite{Morgan79} by time-of-flight measured above 450 keV with better resolution, although he did not provide an analysis of the resonance parameters. Morgan is the reference for the evaluations in the resonance region. More recently, based on the comparison of the aforementioned cross section determined at 123 and 178 keV (at the low-energy tail of the first resonance at 492.7 keV) with the ENDF data in this region Wallner \textit{et al.} suggested a factor of about 3.3 lower strength for this resonance.

The ENDF/B-VIII.0 evaluation does not provide resonance parameters for this reaction, and refers to the Compilation by Ajzenberg-Selove \cite{Ajzenberg-Selove86}. Information on the resonance parameters could be also obtained via other reactions that lead to the same compound nucleus \cite{deBoer20}. Specifically, the inverse \tssc{14}C(p,n)\tssc{14}N reaction, whose threshold allows the observation of the lowest neutron resonances, has been used for providing information on their $J^\pi$ \cite{Bartholomew55,Sanders56,Gibbons59,Niecke77}. Other reactions have thresholds above these states of \tssc{15}N \cite{Ajzenberg-Selove86}. Some measurements found an anisotropy in the first resonance \cite{Sanders56} (indicating $J>\frac{1}{2}$), while others did not \cite{Gibbons59}. 
The neutron polarization measurement by Niecke \textit{et al.} \cite{Niecke77} reported a positive parity for this state. On the contrary, negative parity was assigned in another measurement \cite{Bartholomew55} and compilations \cite{Ajzenberg-Selove86,Mughabghab18}.

To summarize, although the \tssc{14}N(n,p)\tssc{14}C cross-section has been measured several times in different energy ranges, many open questions remain. In order to solve the discrepancies found in the \tssc{14}N(n,p)\tssc{14}C reaction, a measurement covering the range from thermal neutron energy to the first two resonances was carried out at the Experimental Area 2 of the n\_TOF facility at CERN. The neutron beam at this experimental area presents an excellent compromise between high neutron flux, in particular at thermal energy, and energy resolution. The present measurement has covered by far the largest neutron energy range, and in particular, for the first time has connected the thermal, 1/v and resonance regions. This has allowed an accurate analysis of the resonances, providing new information on the cross section in different energy ranges. This measurement is a part of the scientific program of the n\_TOF Collaboration, including a series of experiments aiming at studying reactions of relevant interest in nuclear astrophysics and in medical physics \cite{Chiaveri20}.

\section{EXPERIMENTAL SET-UP}

\subsection{The n\_TOF neutron beam}

The experiment was performed at the Experimental Area 2 (EAR-2) of the n\_TOF facility at CERN, located around 19.5 m from the neutron production target in the vertical direction. The neutrons are generated by the 20 GeV/c proton beam from the CERN Proton Synchrotron (PS) impinging onto a lead target. The technical features of the facility and the characteristics of the neutron beam are described in detail in Refs. \cite{EAR2_design, EAR2_determination}.

\subsection{Detectors and data acquisition}

\begin{figure}[h!]

\centering
\includegraphics[width=.4\textwidth]{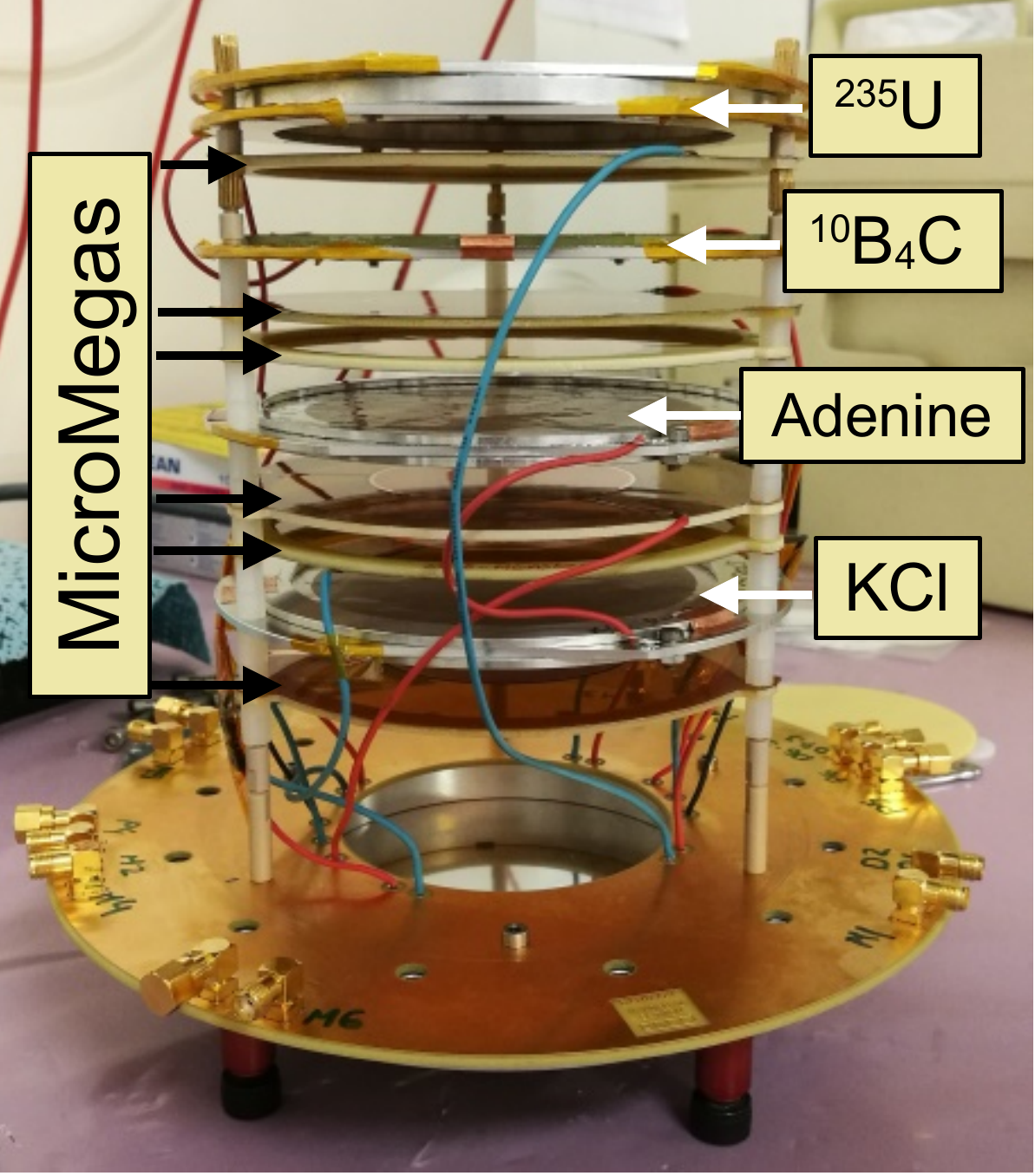} %
\caption{The MicroMegas setup, with the six targets and six MicroMegas detectors used in the measurement. Adenine and KCl samples were placed in back-to-back pairs. }

\label{fig: setupmgas}
\end{figure}

The experimental setup for charged particle detection consisted of a couple of separate systems that worked in parallel. The first of them, (upstream) was based on a stack of micro mesh gaseous structure detectors (MicroMegas), while the second (downstream) used Double-Sided Silicon Strip Detectors (DSSSD).
The MicroMegas detectors are based on microbulk technology, for which the low-mass, robustness and neutron and $\gamma$ transparency allow the use of several detectors along the beam with a minimal perturbation. This allows a geometrical efficiency close to 50 \% when one of the two reaction products are detected \cite{Andriamonje11}. This type of detectors have been used thoroughly in n\_TOF fission and charged-particle emission measurements, including the neutron flux determination \cite{EAR2_determination,Praena18PRC}. A stack of MicroMegas detectors (9.5 cm in diameter) were mounted in a common reaction chamber as shown in Figure \ref{fig: setupmgas}, and operated with a gas mixture of 90 \% Ar and 10 \% CF$_4$. The \tssc{235}U and \tssc{10}B samples were placed in forward direction, the two nitrogen(adenine) samples were placed in a back-to-back configuration, and two KCl samples, which were used for a separate measurement (not analyzed in this work), in the same configuration as the adenine samples. Additional  measurements by replacing the adenine samples with dummy samples (Al foils as in the substrate of the adenine samples) were carried out in order to determine the background and reduce systematic uncertainties.

\begin{figure}[h!]

\centering
\includegraphics[width=.4\textwidth]{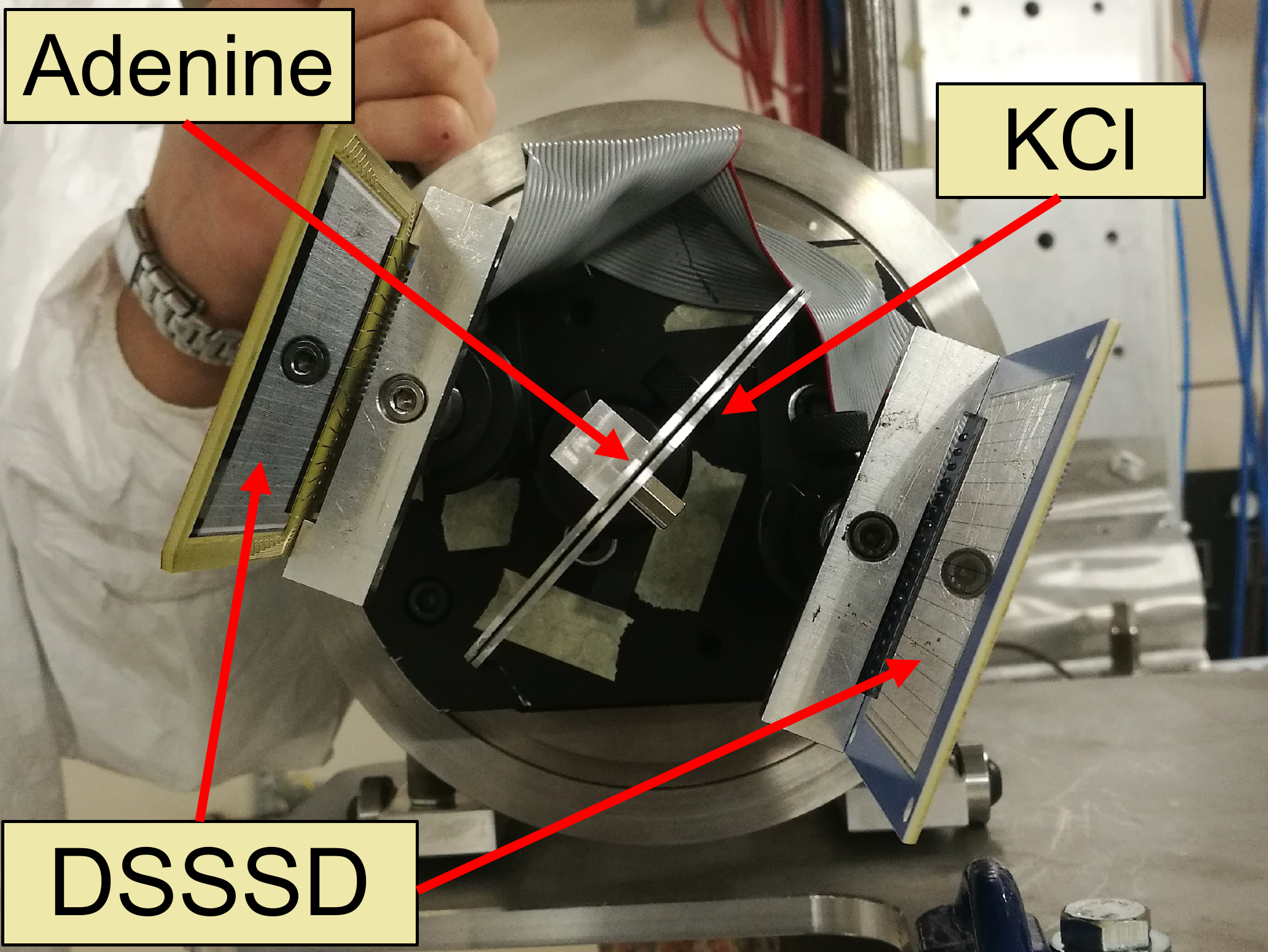}
\includegraphics[width=.4\textwidth]{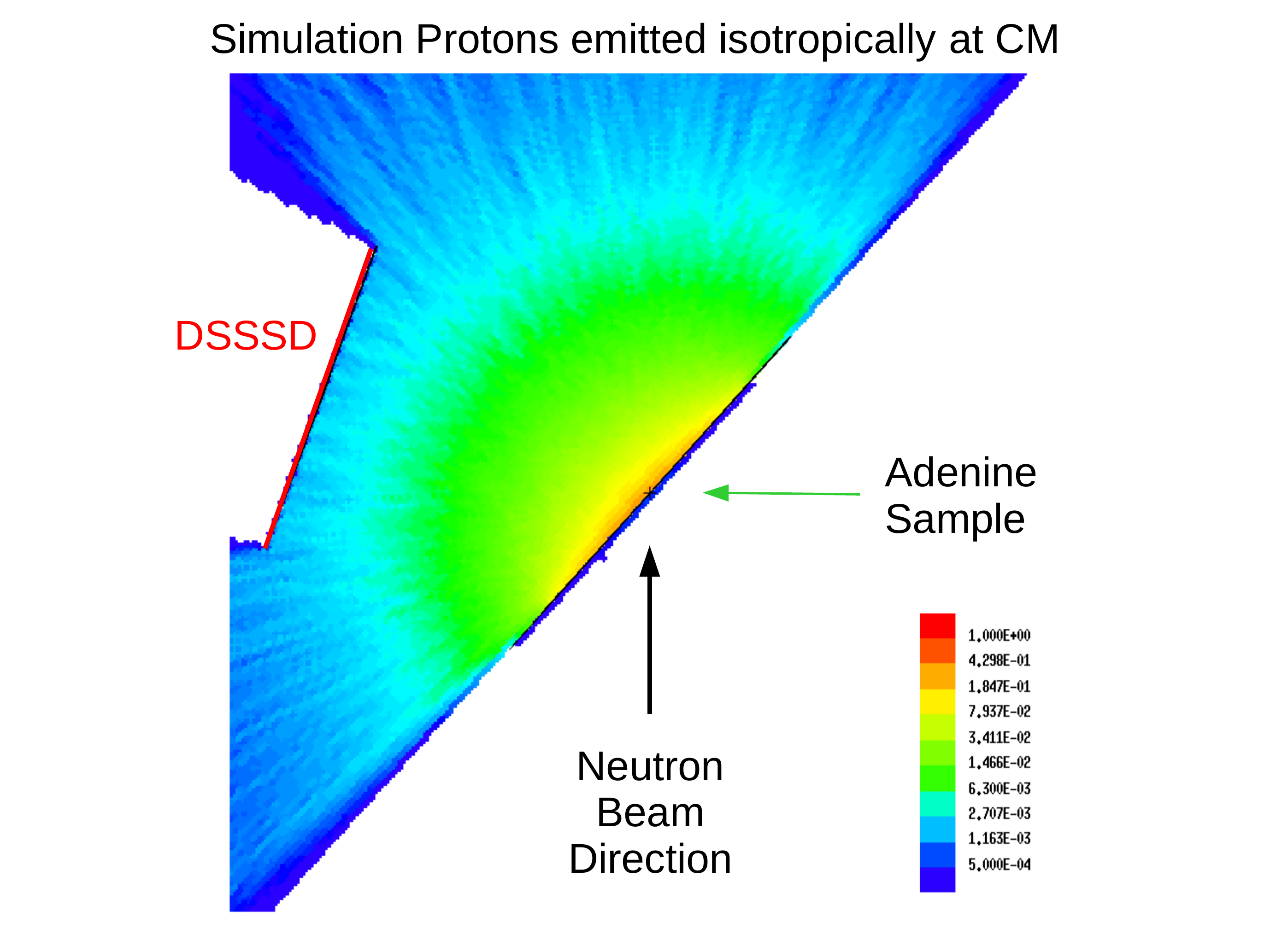}
\caption{\textit{Top:} The DSSSD setup, with the two samples in the center on the Al backings (adenine facing upwards and KCl downwards) and two DSSSD, one facing each sample.
\textit{Bottom:} Simulation to extract the count distribution and detector efficiency considering isotropic proton emission. The position of the adenine sample and the DSSSD are indicated, as well as the incoming neutron beam which serves as the reference direction. The color code corresponds to the proton track density per primary in the simulation.
}

\label{fig: setupDSSSD}
\end{figure}

The DSSSD are silicon detectors that provide position sensitivity and allow background subtraction by means of front-rear strip coincidence analysis. W1 Model DC Strip Detectors with 16$\times$16 strips from Micron Semiconductors were used \cite{DSSSDwebsite}. These detectors have to be placed off-beam where a lower beam-related background is found, at the cost of a lower geometrical efficiency. The DSSSD allow one to extend the measured neutron energy range on the high energy and also analyze the angular distribution of the emitted protons. The setup, shown in Figure \ref{fig: setupDSSSD}, consisted of a couple of detectors, one facing the adenine sample (top) and the other facing the KCl sample (bottom), which was used for a separate measurement as in the MicroMegas setup. The bottom panel of Figure \ref{fig: setupDSSSD} shows a simulation of the protons emitted from the sample after the neutron capture reaction. Protons from the (n,p) reaction impinging on the DSSSD deposited all its energy inside the Silicon layer. Conversely, those protons emitted from the adenine sample in backward direction (likewise for the KCl sample) were absorbed by the Al backings.
These samples were replaced temporarily at the beginning and end of the measurement with \tssc{10}B samples for normalization purposes.

The experimental setup was aligned to the nominal beam position. The real position and spatial profile of the beam was checked by the use of Gafchromic foils. These are radio-sensitive films that contain a dye that changes color when exposed to ionising radiation, providing high resolution on the beam profile distribution. The foils were placed at the position of interest. After irradiation, data was processed through digital scanning. This information was also adopted for corrections in the efficiency related to the beam-to-sample intersection. 

The detector signals were acquired by the standard n\_TOF Data Acquisition System (DAQ), based on SPDevices ADQ412DC-3G cards of 2GS/s maximum sampling rate, 12 bits resolution and 175 MBytes on-board memory \cite{Masi18}. The special features of these cards ensure the collection of data for a time-of-flight (TOF) corresponding to neutron energies well below the thermal energy. 

The signal induced by the prompt $\gamma$-flash generated in the interaction of the 20 GeV/c proton beam with the lead target was used as a reference signal to determine the time of flight of the neutrons.

\subsection{Samples preparation and characterization}

In this measurement, one \tssc{235}U, one \tssc{10}B and two \tssc{14}N samples were used for the MicroMegas setup. For the DSSSD setup, one \tssc{10}B and one \tssc{14}N sample were used. All the sample deposits were 9 cm in diameter in the case of Micromegas, and 5 cm in diameter in the case of DSSSD, which is larger than the neutron beam. The \tssc{235}U sample, enriched to 99.934 \%, was prepared with the electrodeposition method and an areal density of $0.1176\pm0.0005$ mg/cm$^2$ onto 30 $\mu$m of Al. The boron samples, with a thickness of 20 nm (MicroMegas) and 25 nm (DSSSD), were made of \tssc{10}B$_4$C onto Al by the sputtering method. 

The nitrogen samples were made of adenine (C$_5$H$_5$N$_5$) and prepared by thermal evaporation onto Al at CERN. Adenine was used due to a large nitrogen content. At the same time, carbon has a low neutron cross section and neutron capture on hydrogen produces only $\gamma$ radiation that is not detected by the MicroMegas detectors. In addition, the proton recoils from elastic scattering of neutrons have lower energy than those from the \tssc{14}N(n,p) reaction and can be identified and thus filtered out.

The adenine samples were characterized via Rutherford Back-Scattering (RBS) of  $H^+$  with 0.85 MeV at the Centro Nacional de Aceleradores (Spain), where previous works showed excellent possibilities for sample characterization \cite{Praena18}. 
The RBS spectra were analyzed using the \textsc{SIMNRA} package \cite{SIMNRA}. In the SIMNRA simulations the Rutherford cross-section for the scattering of H\tssc{+} in Al was used. For C, N and H, the evaluated cross section data from the IBANDL database were used \cite{IBANDL}. 

\begin{figure}[h!]

\centering
\includegraphics[width=.49\textwidth]{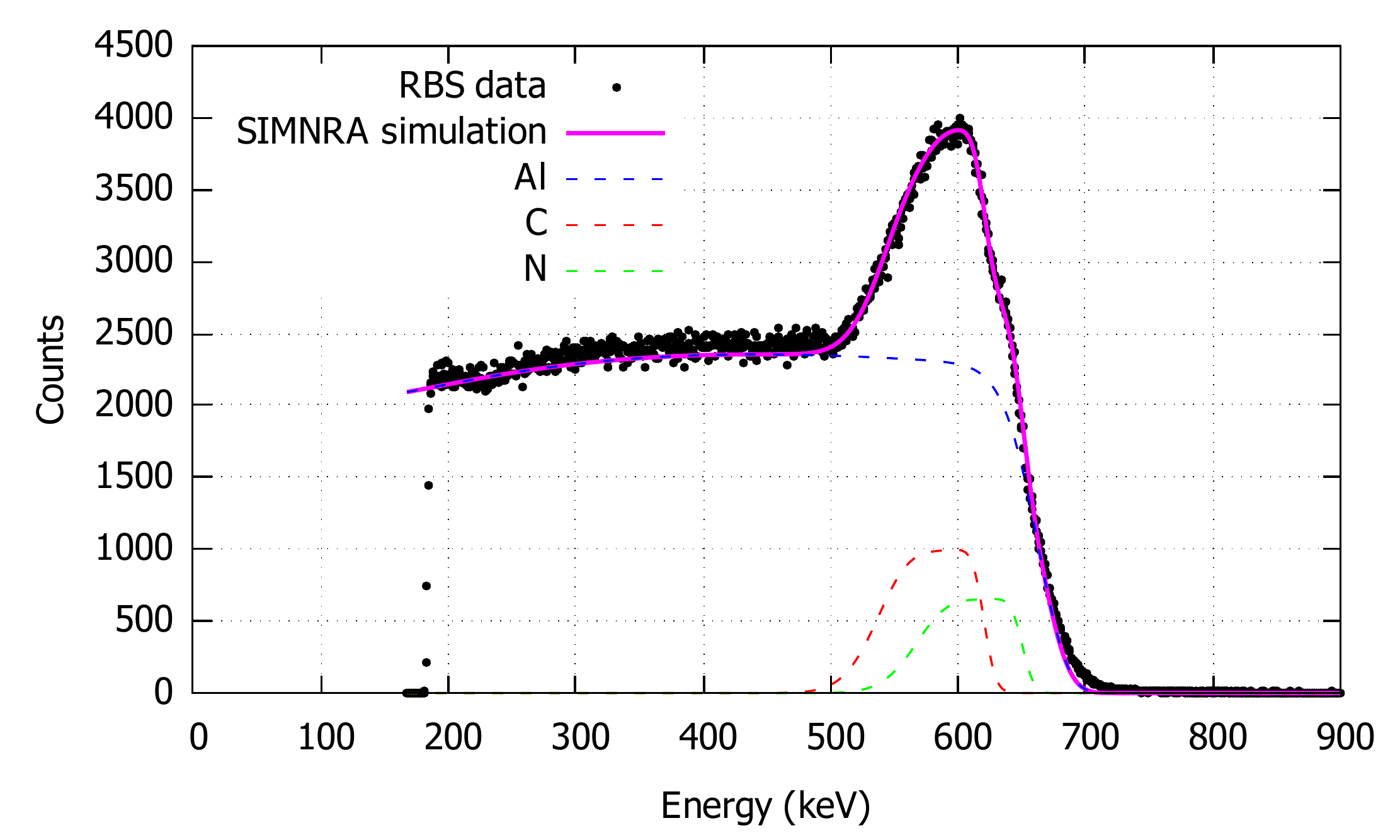}
\caption{ Example of a single RBS measurement. The results of the SIMNRA fit, with the detailed contribution from each isotope is also shown with dashed lines. }

\label{fig: ExpDataSim}
\end{figure}

Considering the dimension of the samples (5 and 9 cm in diameter) and of the H\tssc{+} beam spot (3 mm) used for the characterization of the samples, several points were analyzed for each sample to provide a picture of the homogeneity of the thickness. The samples were scanned from the edges to the center in three different directions. In order to perform an accurate and precise determination of the number of atoms of \tssc{14}N, few points outside the area coated with adenine were also analyzed by RBS. This allowed the determination of any possible contamination of the substrate reducing the free parameters of the SIMNRA fit of the data. The overall thickness of adenine is revealed both by the presence of the peak and the reduction in energy of the edge from the Al substrate. Figure \ref{fig: ExpDataSim} shows an example of SIMNRA fit to the data. The mass density for each measured point was determined with uncertainty of 1-2 \%.

\begin{figure}[h!]

\centering
\includegraphics[width=.49\textwidth]{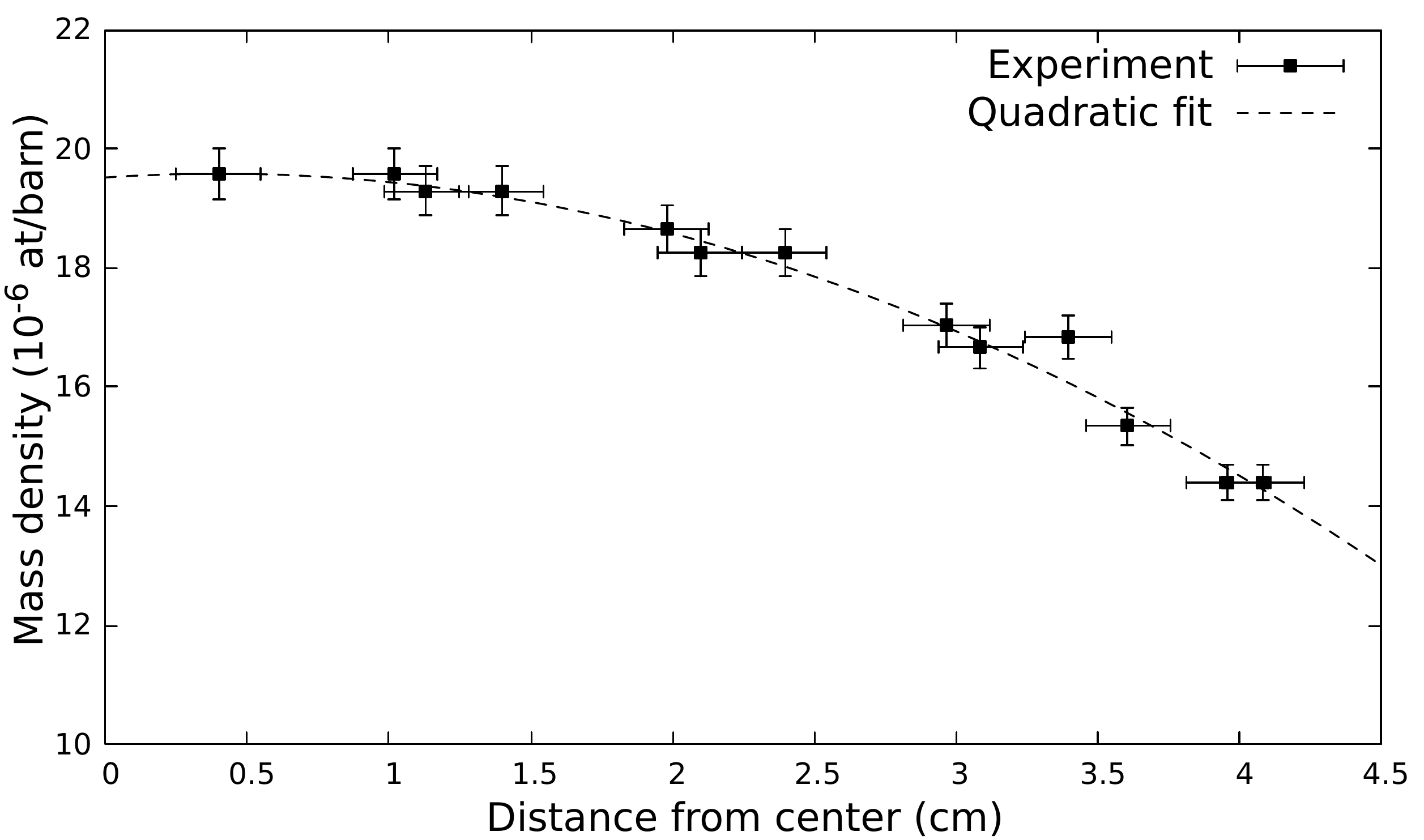}
\caption{Mass density radial distribution for the Forward adenine sample in the MicroMegas setup. The horizontal error bars match with the size of the probe (3 mm). The quadratic fit is shown with a dashed line. }

\label{fig: RadialDistr}
\end{figure}

 We found a smooth reduction of the adenine mass density from the center to the edges. Within uncertainties, the same mass density was found for points at the same distance from the center.  Figure \ref{fig: RadialDistr} reveals a parabolic pattern in the mass density, which was attained through the equation  $m = m_{0}- a\cdot r^2$. The parameter $a$ is a measure of the mass distribution throughout the sample, quantified as the curvature of the quadratic fit, while $m_{0}$ is the mass density at the center of the samples. The total mass is computed as the integral of the mass density given by the above mentioned formula. The mass of each sample was determined with an uncertainty better than 1.5 \%. Table \ref{tab:tablemasses} summarizes the results for all the samples. 

\begin{table*}[h!]
    \centering
        \begin{tabular}{ccccc}
        \hline\hline
        Sample & $m_{0}$ & $a$  & Total number of atoms   \\ 
        & ( at/b) & (at/b/cm\tssc{2}) & ($10^{20} \cdot $at) 
        \\
        \hline
        DSSSD             & 6.88$\cdot10^{-6}$   & 7.24$\cdot10^{-8}$     & 1.306$\pm$0.020    \\
        Backward MicroMegas & 1.382$\cdot10^{-5}$     & 2.18$\cdot10^{-7}$    & 8.79$\pm$0.11    \\
        Forward MicroMegas  & 1.958$\cdot10^{-5}$    & 3.91$\cdot10^{-7}$     & 9.94$\pm$0.12   \\ \hline \hline
\end{tabular}
    \caption{Results of the characterization of the mass density for all the adenine samples.}
    \label{tab:tablemasses}
\end{table*}

\section{DATA ANALYSIS}
\label{sec:DataAnalysis}

The digitized signals from MicroMegas detectors and DSSSD were reconstructed off-line by means of a Pulse Shape Analysis routine described in Ref. \cite{Zugec16}, from which information was extracted on the amplitude, area, timing and other features of the signals. The analysis was done separately for high intensity (HI) and low intensity (LI) proton pulses from the PS accelerator complex. High intensity pulses (around 7·10\tssc{12} protons per bunch) allow larger statistics, but suffer from a very intense $\gamma$-flash. On the other hand, low intensity pulses (around 3.5·10\tssc{12} protons per bunch) profit from a reduced $\gamma$-flash which allows a better signal identification, specially for higher neutron energies. The use of LI pulses thus extends the energy range of the measurement in the high energy region, at the expense of a smaller neutron flux and hence lower statistics.

\begin{figure}[h!]

\centering
\includegraphics[width=.49\textwidth]{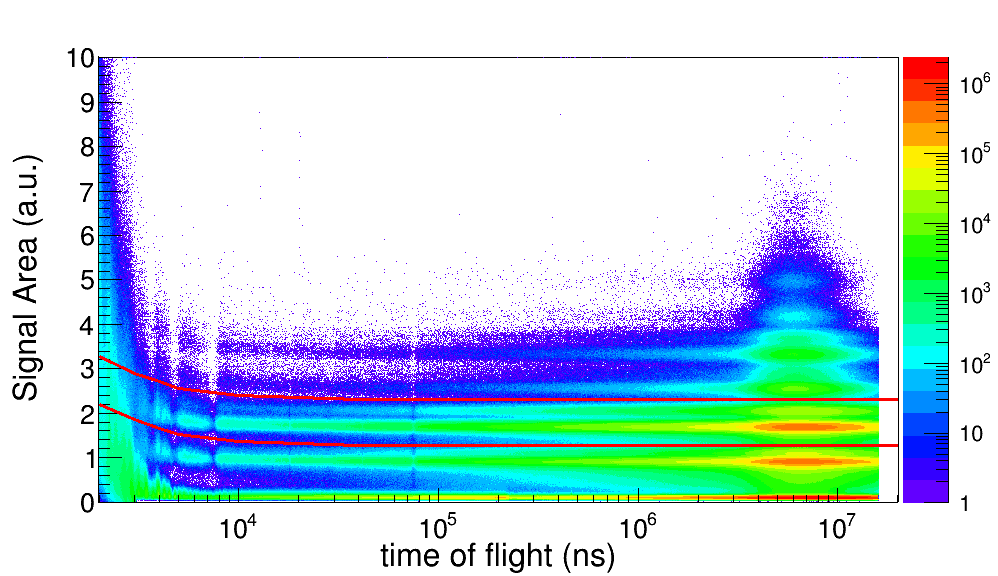}
\caption{2D histogram of deposited energy vs. time of flight for the MicroMegas detector in forward emission from \tssc{10}B. The lines indicate the energy-dependent range applied to select the $\alpha$-particles used in the analysis. The two regions corresponding to the detection of $\alpha$-particles and \tssc{7}Li are clearly distinguished above and below the bottom red line, respectively.  Pile-up events can also be observed above the regular signals from $\alpha$ particles. The behavior at low TOF is related to the effect of the $\gamma$-flash.
}

\label{fig:borontofarea}
\end{figure}

Figure \ref{fig:borontofarea} shows a 2D histogram of the area of the signals versus their TOF for the MicroMegas detector facing the \tssc{10}B sample, where signals from $\alpha$ particles are discriminated from the other reaction products, electronic noise and pile-up events. The $\alpha$-particles from the two reaction channels: the decay to the ground state (n, $\alpha_0$) and to the first excited state (n, $\alpha_1 \gamma$) can be observed around 2.0 and 1.7 in area, respectively. Some regions with low number of counts are observed, specially below 10\tssc{4} ns, which correspond to dips in the neutron flux caused by Al in the target and beam pipes. 

\begin{figure}[h!]

\centering
\includegraphics[width=.49\textwidth]{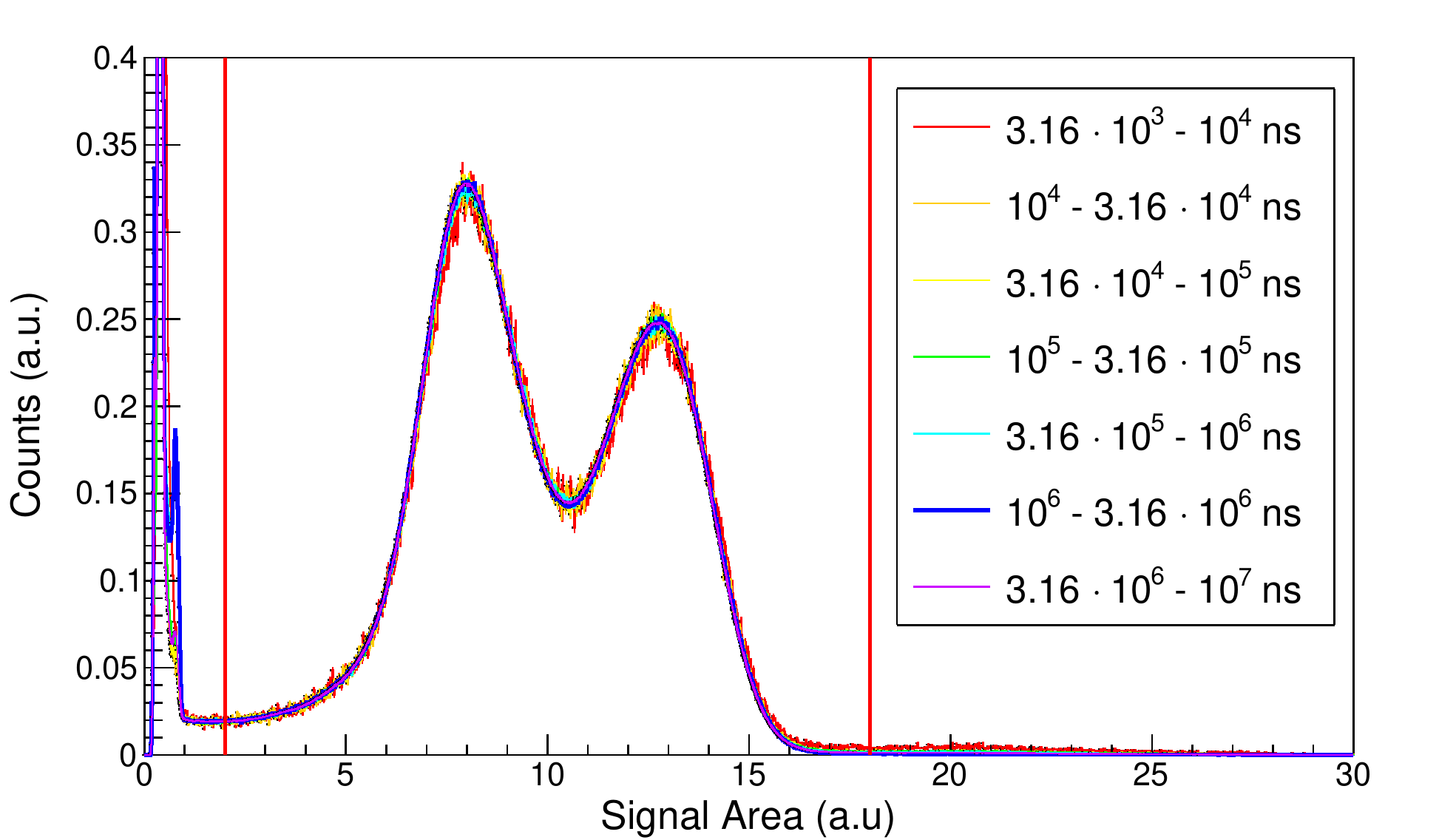}
\caption{Energy deposition spectra for seven TOF intervals at the MicroMegas detector facing the \tssc{235}U sample. The two-bump structure corresponds to the fission fragments asymmetric masses. The curves have been normalized to the same integral between 2 and 18 and indicate that the fraction of fission fragments lost below the threshold is independent of the neutron energy.
}

\label{fig:areayieldu}
\end{figure}

Similarly, a threshold in energy deposition has been used to separate the fission fragments from the signals from $\alpha$ decay and electronic noise for the detector facing the \tssc{235}U sample as shown in Figure \ref{fig:areayieldu}. The Figure shows the projection of the signal areas on the vertical axis (of a plot similar to Figure \ref{fig:borontofarea}) for several TOF ranges. Individual curves were normalized to prove that the distribution is the same at all energies, hence the correction due to the fraction of fission events lost below the threshold can be considered to be the same at all TOF values.

The residual background within the selection thresholds was measured by means of dummy samples. The relative background at all the neutron energies was found well below $10^{-3}$ for the \tssc{235}U and \tssc{10}B samples, therefore it was neglected in the analysis.

\begin{figure}[h!]

\centering
\includegraphics[width=.49\textwidth]{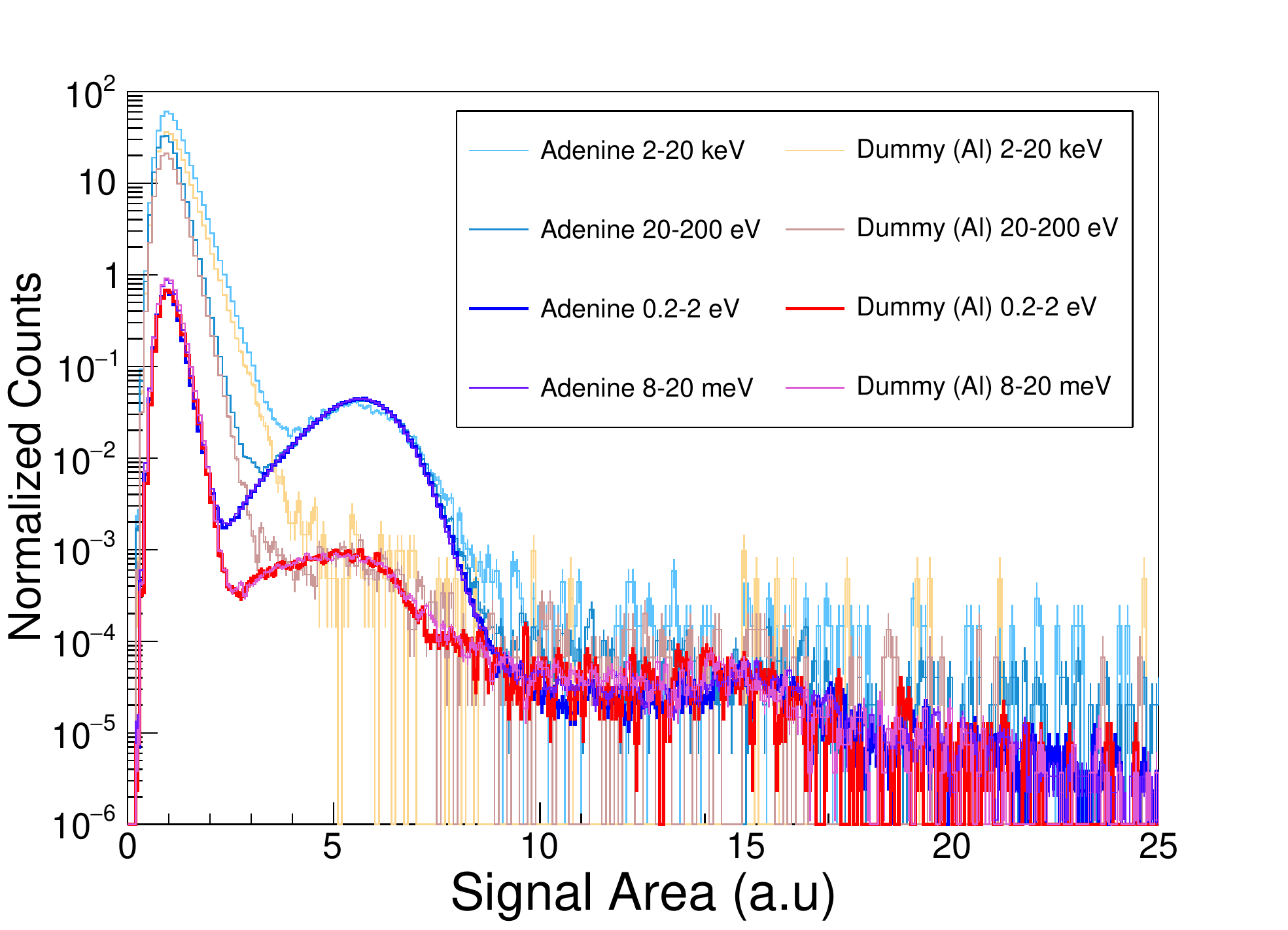}
\caption{Energy deposition spectra for four neutron energy intervals  in the MicroMegas detector facing the forward adenine sample. The adenine spectra have been normalized to the same integral between 3.7 and 10. The background spectra were normalized to the adenine ones using the total proton intensity from the PS pulses.
}

\label{fig:N3MGASarea}
\end{figure}

In the case of the adenine samples, the emitted proton has a lower energy than the $\alpha$ and \tssc{7}Li from \tssc{10}B capture, which makes it more complicated to separate their signals from the low  signal area noise and background. Fittings of the count distribution along the whole neutron energy range were performed, in order to estimate the fraction of rejected counts below the selection threshold. Figure \ref{fig:N3MGASarea} shows the counts from the forward adenine sample and the dummy sample used to estimate the background for several neutron energy ranges.  The signals from protons between 3.7 and 10 in area can be clearly distinguished above the background. Note an increase of the noise (signals with area lower than about 3) in the detector at higher neutron energies. This noise limits the precision of the background subtraction at neutron energies above 130 keV. Note also the slight increase in energy deposition at higher neutron energies, due to kinematics.
\label{sec:adeninesamples}

\begin{figure}[h!]

\centering
\includegraphics[width=.49\textwidth]{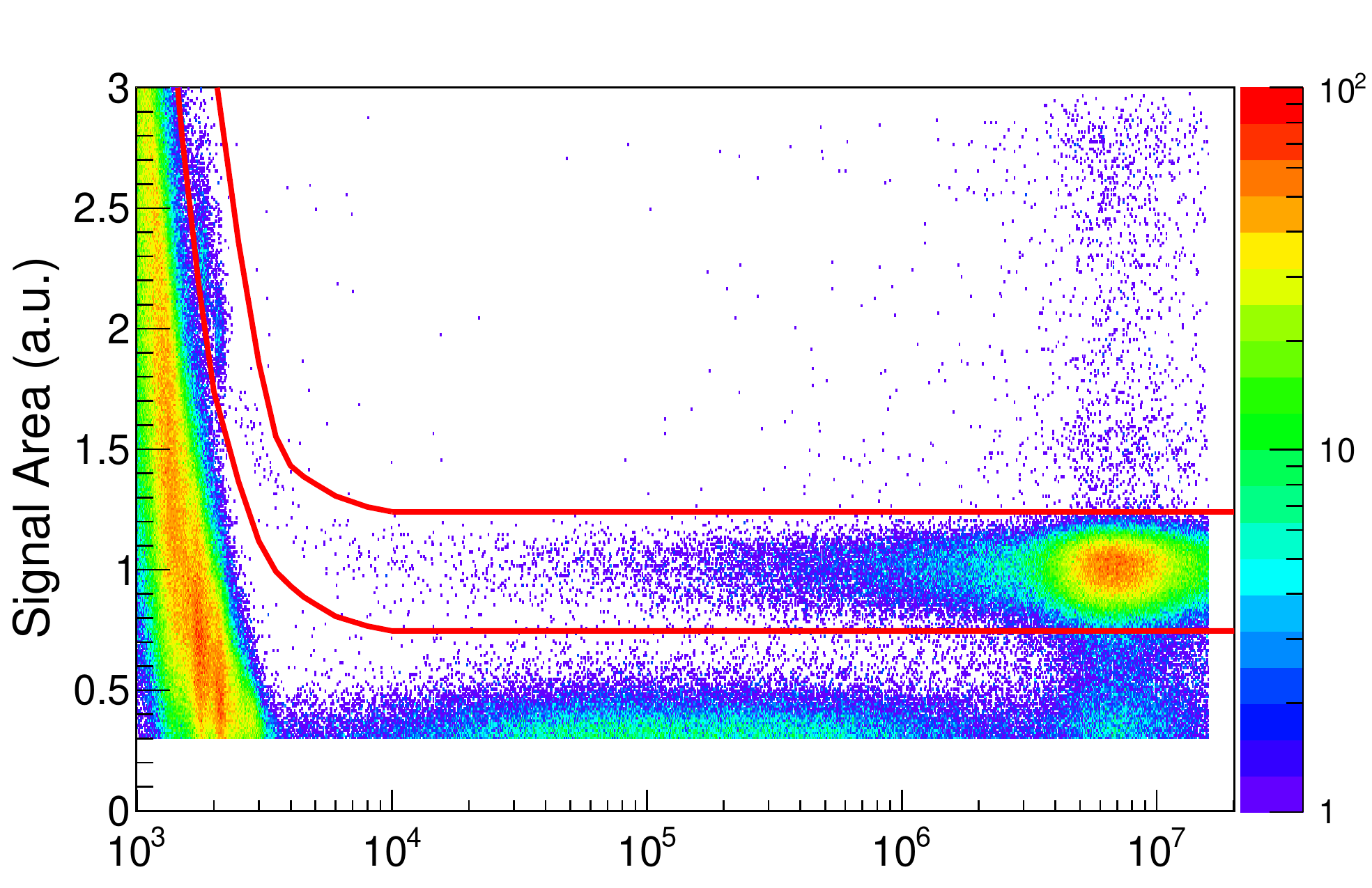}
\includegraphics[width=.49\textwidth]{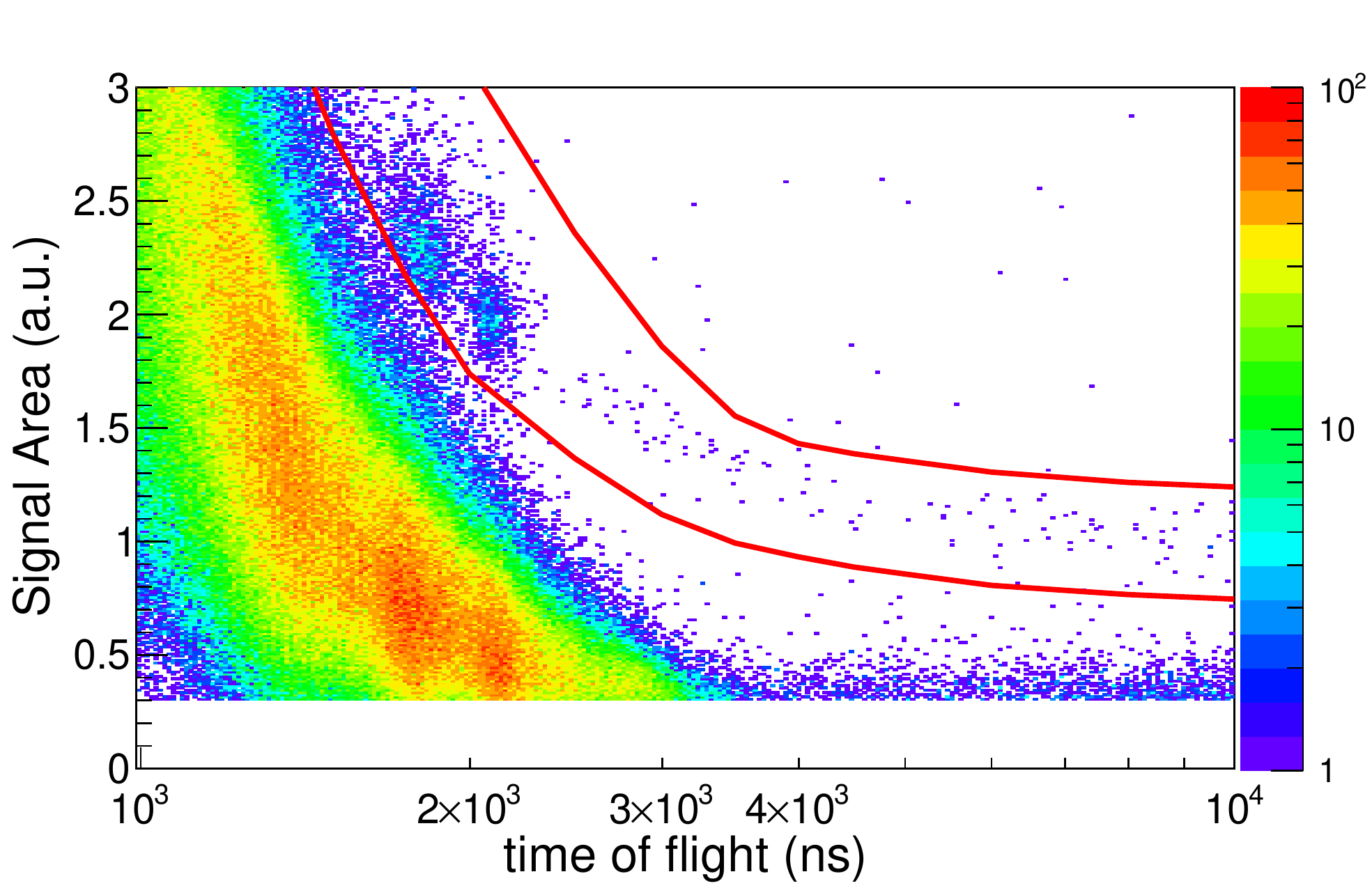}
\caption{2D histogram of deposited energy (signal area) vs. TOF for the DSSSD facing the adenine sample. The spectra correspond only to the LI pulses. The red lines indicate the region to select the protons from the \tssc{14}N(n,p) reaction used in the analysis. Signals below 0.3 in signal area are not shown as they are strongly dominated by the electronic noise. The signals corresponding to the first two resonances can be clearly identified around 2000 ns. The huge number of events at short TOF (lower than 3·10\tssc{3} ns) is a consequence of the $\gamma$-flash. The bottom panel is a zoom in the low TOF range.
}

\label{fig:NDSSSDtofarea}
\end{figure}

For the analysis of measurement with DSSSD, only coincidence signals from the front and rear strips in the detector were considered in order to reduce part of the background signals appearing at high neutron energies. A time and signal area coincidence rejection was performed. Figure \ref{fig:NDSSSDtofarea} shows a 2D histogram of the signal area versus their TOF for the DSSSD facing the adenine sample. The use of only coincident signals has significant impact specially at TOF corresponding to the first two resonances, where we already see impact of the $\gamma$-flash. Only signals within a neutron energy-dependent signal area region, enclosed in red lines in Figure \ref{fig:NDSSSDtofarea}, were considered in the analysis. 
A small background was subtracted using collected data from dummy samples (Al backings). Nonetheless, there was an additional small background contribution that can not be easily subtracted, which was considered in the R-matrix (\textsc{sammy}) analysis of the data. Pile-up events are negligible for all neutron energies due to the low counting rate, given the off-beam location of the DSSSD. A small amount of boron contamination is observed at large TOF from\tssc{7}Li and $\alpha$ signals. They have higher signal area than the protons (around 1.5 and 2.8 in area in the plot, respectively), which hence do not affect the analysis. As mentioned above, the spectra from LI proton pulses allow analysis up to a higher energy as the background is lower than for the HI ones. Both HI and LI data were used below 300 keV and only LI data above that energy, resulting in lower collected statistics in the resonance region. 

\subsection{Energy calibration}

\begin{figure}[h!]

\centering
\includegraphics[width=.49\textwidth]{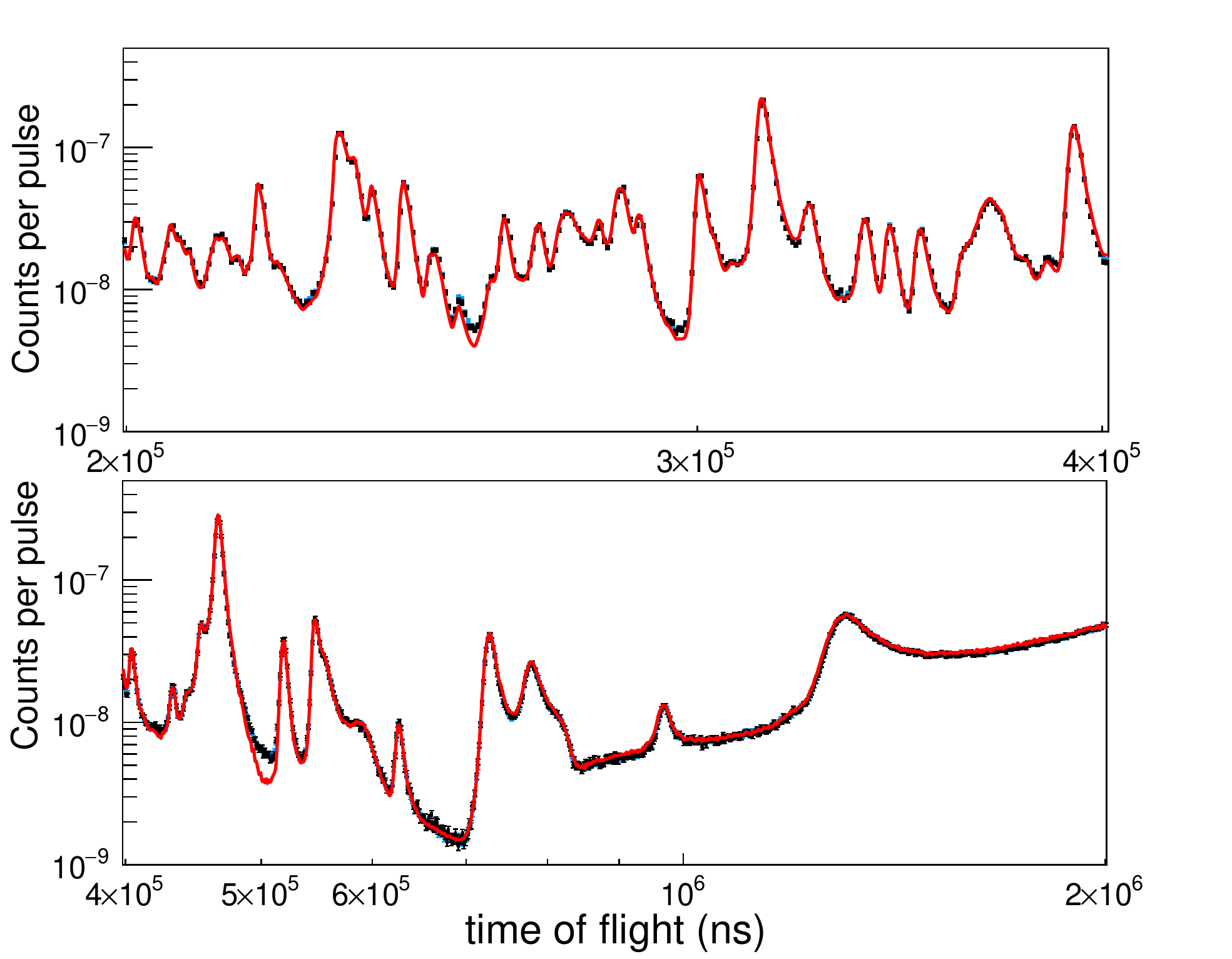}
\caption{Measured \tssc{235}U TOF spectra with MicroMegas in the low energy and resonance
region (blue points for LI pulses, black points for HI pulses), compared to the ENDF/B-VIII.0
evaluation (red line) convoluted with the n\_TOF EAR-2 resolution function by means of the n\_TOF Transport Code. Small differences observed in some valleys come likely from improper values in ENDF/B-VIII.0 - compare this database \textit{e.g.} with JEFF-3.3.
}

\label{fig:tofyieldu}
\end{figure}

The TOF-to-energy calibration was performed according to the method described in Ref. \cite{RFtechreport}. The experimental TOF yield was compared to the simulations from the n\_TOF Transport Code, which includes ENDF/B-VIII.0 cross-section evaluation with the effect of the n\_TOF EAR-2 resolution function. This comparison is shown for the low energy resonances of \tssc{235}U in Figure \ref{fig:tofyieldu}. There is an overall good reproduction in the whole range, except in the dips between some of the resonances, which had already been spotted as disagreements between the last ENDF/B-VIII.0 and JEFF-3.3 evaluations.
The extracted effective flight path was 19.39 m for the \tssc{235}U sample position in the MicroMegas Chamber. This corresponds to the geometrical distance of the experimental apparatus from the surface of the Pb spallation target. This flight path was adjusted for the subsequent samples according to their position inside the Micromegas chamber. For the DSSSD setup, the flight path was 19.75 m and it was checked by the fitting of the thermal peak and the dips in the flux observed with the reference \tssc{10}B sample.


\subsection{Neutron Fluence and normalization}

The \tssc{14}N(n,p) cross section is extracted relative to the \tssc{10}B(n,$\alpha$) cross section according to:

\begin{equation}
    \sigma_{N}(E_n) = \sigma_{B}(E_n) \frac{\left(C_{N}(E_n)-B_{N}(E_n)\right)\Phi_{B}(E_n) N_{B}\epsilon_{B}}{\left(C_{B}(E_n)-B_{B}(E_n)\right)\Phi_{N}(E_n) N_{N}\epsilon_{N}},
    \label{eq:xscalc}
\end{equation}

where $C_X$ ($X\equiv$ N or B for nitrogen and boron, respectively) is the total number of counts in the detector at a given neutron energy $E_n$ and $B_X$ is that of background counts, $\Phi_X$ is the neutron fluence, $N_X$ is the areal density of the sample (in atoms per barn) and $\epsilon_X$ is the efficiency, which accounts for the geometric efficiency and the event selection thresholds. Reference boron cross-section data was taken from Ref. \cite{Carlson18}.

\begin{figure}[h!]

\centering
\includegraphics[width=.49\textwidth]{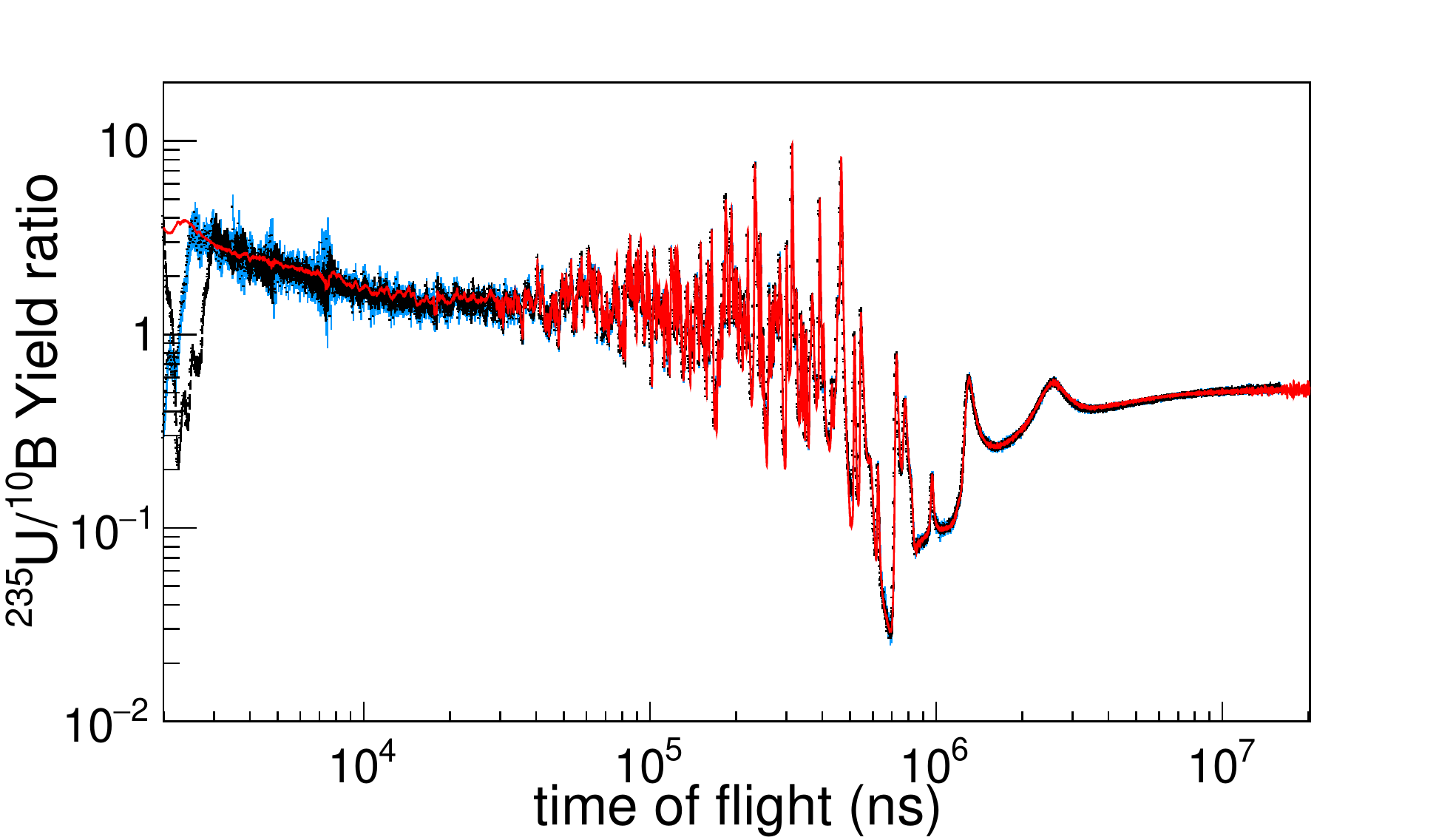}
\caption{Measured \tssc{235}U/\tssc{10}B yield ratio for HI (black) and LI (blue) PS proton pulses, compared to the ENDF/B-VIII.0 evaluation (red) folded with the n\_TOF EAR-2 resolution function. }

\label{fig:ubratio}
\end{figure}

Along with this, the \tssc{235}U/\tssc{10}B yield ratio was calculated for validation purposes. A good agreement is found, as shown in Figure \ref{fig:ubratio}, comparing the ratio of the TOF count spectra to the simulations, where the different detection efficiencies and the neutron beam resolution function are considered, following Ref. \cite{RFtechreport}. The effect of neutron flux attenuation due to the relative position of the uranium and boron samples was also considered.

\subsection{Efficiency and dead-time corrections}
\begin{figure}[h!]

\centering
\includegraphics[width=.49\textwidth]{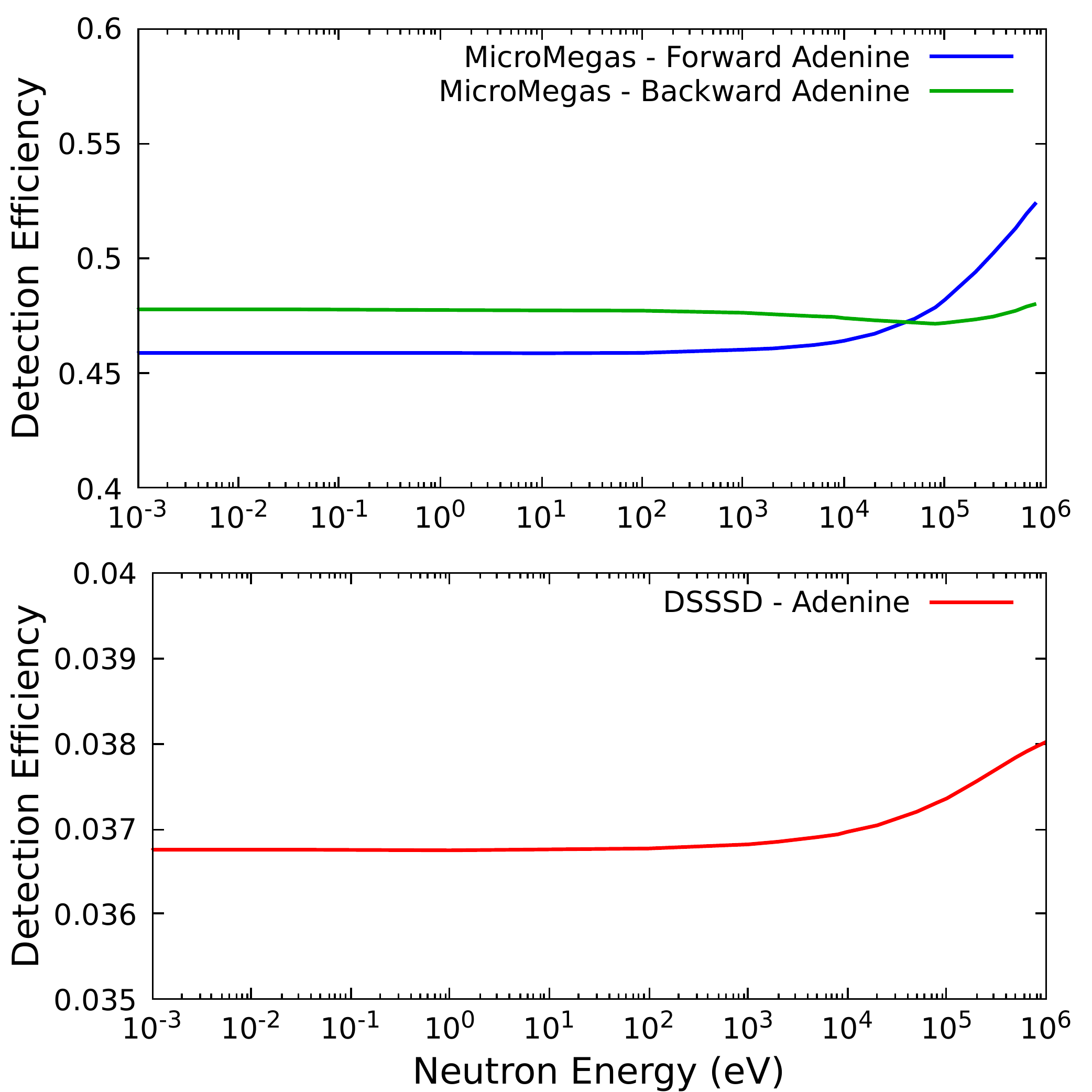}
\caption{Upper panel: Simulated Detection Efficiency of the MicroMegas Detectors. 
Lower panel: Simulated Detection Efficiency of the DSSSD.
Note the different vertical scale for Micromegas and DSSSD.
}

\label{fig:effsim}
\end{figure}

The efficiency corrections for the Micromegas detectors were determined by detailed Monte Carlo simulations of the reaction products' energy loss in the samples and the gas. The simulations were performed with the MCNP6.2 code \cite{MCNP62}. In these simulations, the actual thickness profiles that were found experimentally have been used. Energy and angular distributions of emitted charged-particles are adopted from \cite{Hambsch09} for the boron sample, and is assumed as isotropic at the center of mass for U and for adenine samples at low energy. See Sec. \ref{sec:angulardistr} for the check of the distribution from N in the resonance region. The reaction events were assumed to occur uniformly along the beam direction inside the sample, and the beam profile from the n\_TOF Transport Code was used.  Figure \ref{fig:effsim} shows the efficiency for the forward and backward adenine samples. As evident from the Figure, the efficiency changes only weakly. The change is seen (due to kinematics) when the incident neutron energy becomes comparable to the reaction Q-value. The efficiency at low neutron energy is different for forward and backward samples due to the different thickness of the samples (larger for the forward sample), which becomes relevant for proton emission near grazing angles, specially at higher energies. Furthermore, the increasing linear momentum transfer contributes to a larger emission in the forward direction and thus a higher efficiency in the forward case.

For the DSSSD setup, the geometry of the detectors was implemented in the simulations, including the silicon layer thickness and the double strip features, specially regarding the inter-strip spacing. The positioning and orientation of the detector with respect to the sample and the beam were checked by using an \tssc{241}Am source and also the counts from the thermal neutron capture on \tssc{10}B, assuming isotropic $\alpha$ particle emission. Furthermore, some of the strips from the DSSSD were malfunctioning and have been kept out of the analysis. The simulated detection efficiency for the DSSSD adenine sample is shown in the lower panel of Figure \ref{fig:effsim}.

Dead-time corrections were computed following the non-paralyzable response model \cite{Moore80} assuming a fixed dead time for each detector. The dead times were estimated following two strategies. The first was based on estimating the dead-time from the FWHM of the signals and the minimum time difference between consecutive pulses (a sharp cut-off observed in the time-between-pulses distribution). The second method relied on matching the total yield for low and high intensity proton pulses, showing agreement with the first method.

\subsection{Uncertainties}

We adopted conservative estimations of the accuracy of the results. The statistical uncertainty of the MicroMegas data was assessed with the use of a binning of 10 bins per decade (bpd), ranging from 0.1 \% around the thermal point to the 4-5 \% above 10 keV. Statistical uncertainties of the DSSSD data, at 10 bpd, ranges from 0.4 \% at thermal, to 10-15 \% in the 10-300 keV range. Above 300 keV, the need of a better energy resolution due to the presence of resonances makes us use 100 bpd, at a cost of a higher uncertainty close to 20 \% per bin.

The mass of the adenine samples was measured within 1.5 \% accuracy for the MicroMegas samples.  Although the neutron energy dependence of the \tssc{14}N(n,p) cross-section was determined with respect to \tssc{10}B at all neutron energies (as indicated by Eq. \ref{eq:xscalc}), the absolute cross-section normalization was made using the thermal point of the \tssc{235}U(n,f) reaction, given that it is a standard at that energy and the U sample mass was known more precisely, with uncertainty of 0.43 \%, than the B mass. The uncertainty of the cross-section of \tssc{235}U(n$_{th}$,f) reaction is 0.23 \% \cite{Carlson18}. 

The uncertainty related to the angular distribution of the $\alpha$ particles from the \tssc{10}B(n,$\alpha$)\tssc{7}Li reaction at higher energies was considered in the calculation of the efficiency in both the MicroMegas and DSSSD setups following the accurate data in Ref. \cite{Hambsch09}. The uncertainty due to the detection efficiency includes the statistical uncertainty in the simulations of the proton transport (and $\alpha$ particles for the \tssc{10}B${_4}$C sample), which was reduced below 0.2 \%. Additional uncertainty due to the positioning and orientation of the samples and DSSSD was estimated by simulations with perturbations on the geometry to be 1.3 \%. The uncertainty in the correction to the selection cuts was estimated to be 1 \% at thermal energy and up to 5 \% in the keV range. The effect of the neutron beam attenuation and neutron scattering on Al windows in the n\_TOF pipes, Kapton windows of the MicroMegas and DSSSD chambers, the in-beam Micromegas detectors and upstream samples was assessed via simulations and found around 1 \%. Since DSSSD results are normalized to the MicroMegas data at the 1/v region, additional systematic uncertainties from this detection system can be dismissed.  The sources of uncertainty are summarized in Table \ref{tab:tableuncertainties}.

\begin{table*}[h]
    \centering
        \begin{tabular}{lc}
        \hline\hline
        Component & Uncertainty (\%) \\ \hline
        Sample mass \tssc{14}N    & 1.2-1.5 \\
        Normalization & 0.65-3 \\
        Efficiency and selection cuts & 1.8-5 \\
        Neutron Beam attenuation & 1 \\
        Statistical MGAS & 0.1-5\\
        Statistical DSSSD & 0.4-20\\

        \hline\hline
\end{tabular}
    \caption{Major sources of uncertainty (in \%) of the present \tssc{14}N(n,p) cross section. The uncertainties vary depending on the binning and the energy range and on the sample. }
    \label{tab:tableuncertainties}
\end{table*}

\section{RESULTS}
\label{sec:results}
\subsection{Data from MicroMegas and DSSSD}

The main goal of this measurement was to obtain a consistent data set spanning from thermal to the resonance region and to provide a verification of the first two resonances strengths after a recent paper \cite{Wallner16} suggested a possible reduction by a factor of 3.3.

The measurement with MicroMegas detectors has covered the range from 8 meV to 80 keV with the backward sample, and up to 130 keV with the forward sample. The extended range for the forward sample is due to its higher mass and slightly higher efficiency at these energies, which leads to an improved signal-to-background ratio. The results from both samples agree with each other within uncertainties. The measurement covers the 1/v range, including the thermal point, reproducing and extending the data by Koehler \cite{Koehler89}. The high-quality measurement with MicroMegas detectors profits from thick samples and a large detection efficiency.

The measurement with DSSSD detectors then allows to extend the range from thermal to 800 keV. This allows to fully cover the astrophysical range of interest, and allows to determine the parameters of the first two observed neutron resonances. The data from DSSSD have a lower counting statistics due to a smaller thickness of the sample and lower geometrical efficiency.  The DSSSD data have been normalized to that of MicroMegas detectors in the range 8 meV - 100 eV, where there are data from both detection systems.

\begin{figure}[h!]

\centering
\includegraphics[width=.49\textwidth]{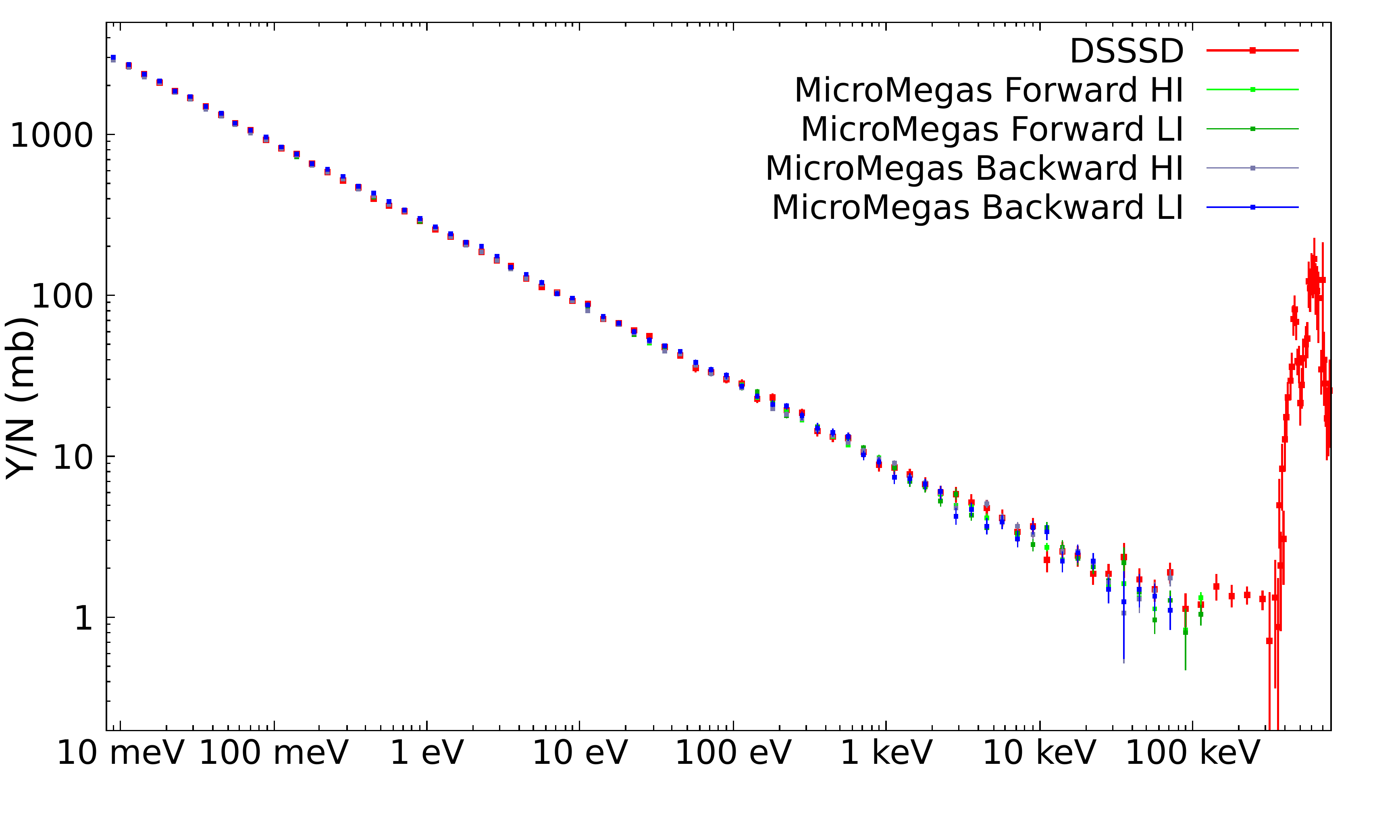}
\includegraphics[width=.49\textwidth]{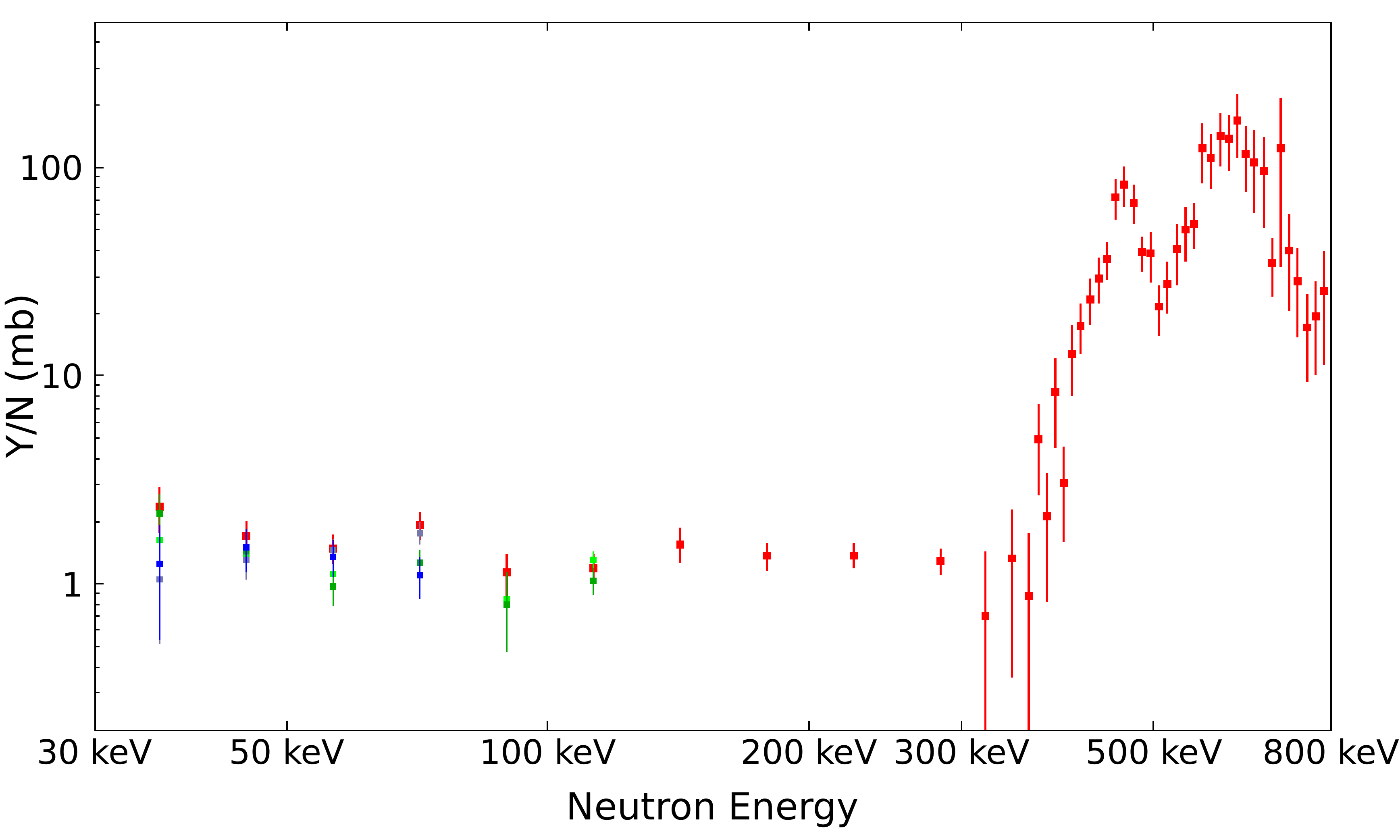}
\caption{\tssc{14}N(n,p) Yield/Areal Density ($Y/N$) from the experiment. Data are shown at 10 bins per decade below 300 keV and 100 bins per decade above 300 keV. MicroMegas results are separated for forward and backward samples, and also in HI and LI PS proton pulses. 
}
\label{fig:N14Data}
\end{figure}

Overall, this measurement spans eight orders of magnitude of neutron energy providing for the first time a common consistent dataset for the thermal and 1/v range, the astrophysical range and the resonance region. Figure \ref{fig:N14Data} shows the experimental yield corrected for areal density ($Y/N$). This quantity differs from the cross-section only by the effect of the n\_TOF EAR-2 resolution function. Data is presented for both MicroMegas detectors (separately for the forward and backward samples and HI and LI proton pulses) and DSSSD (merged HI and LI below 300 keV, and only LI above 300 keV). 

The cross-section derived from this yield by means of its fitting with the R-Matrix based \textsc{sammy} code \cite{SAMMY} will be shown below. Before this, the angular distribution of the proton emission in the first resonance will be discussed.

\subsection{Angular distribution of protons from the 492.7 keV resonance}
\label{sec:angulardistr}

Previous measurements of the \tssc{14}N(n,p) reaction did not carry out any analysis of angular distribution of protons emitted from the resonances. As the DSSSD detects protons emitted only in a range of angles, the knowledge of the angular distribution of emitted protons is crucial for the cross-section determination. In practice, the DSSSD allows to check the distribution for the first time, even considering its moderate angular resolution. Any deviation from isotropy would be an indication that the spin of these resonances is  $J>\frac{1}{2}$ and would complicate the determination of the cross-section, given the relevance of the angular distribution in the efficiency calculation of the DSSSD set-up.
The spin and parity for the corresponding \tssc{15}N state was assigned as $\frac{1}{2}^{-}$ from a measurement of the \tssc{14}C(p,$\gamma$) reaction \cite{Bartholomew55}.  Previous measurements of the inverse reaction \tssc{14}C(p,n)\tssc{14}C then determined the angular distribution of neutron emission, with Sanders \cite{Sanders56} observing a possible anisotropy but attributing it to target non-uniformity, and later Gibbons \textit{et al.} not observing any anisotropy at the center of mass \cite{Gibbons59}.

\begin{figure}[h!]

\centering
\includegraphics[width=.49\textwidth]{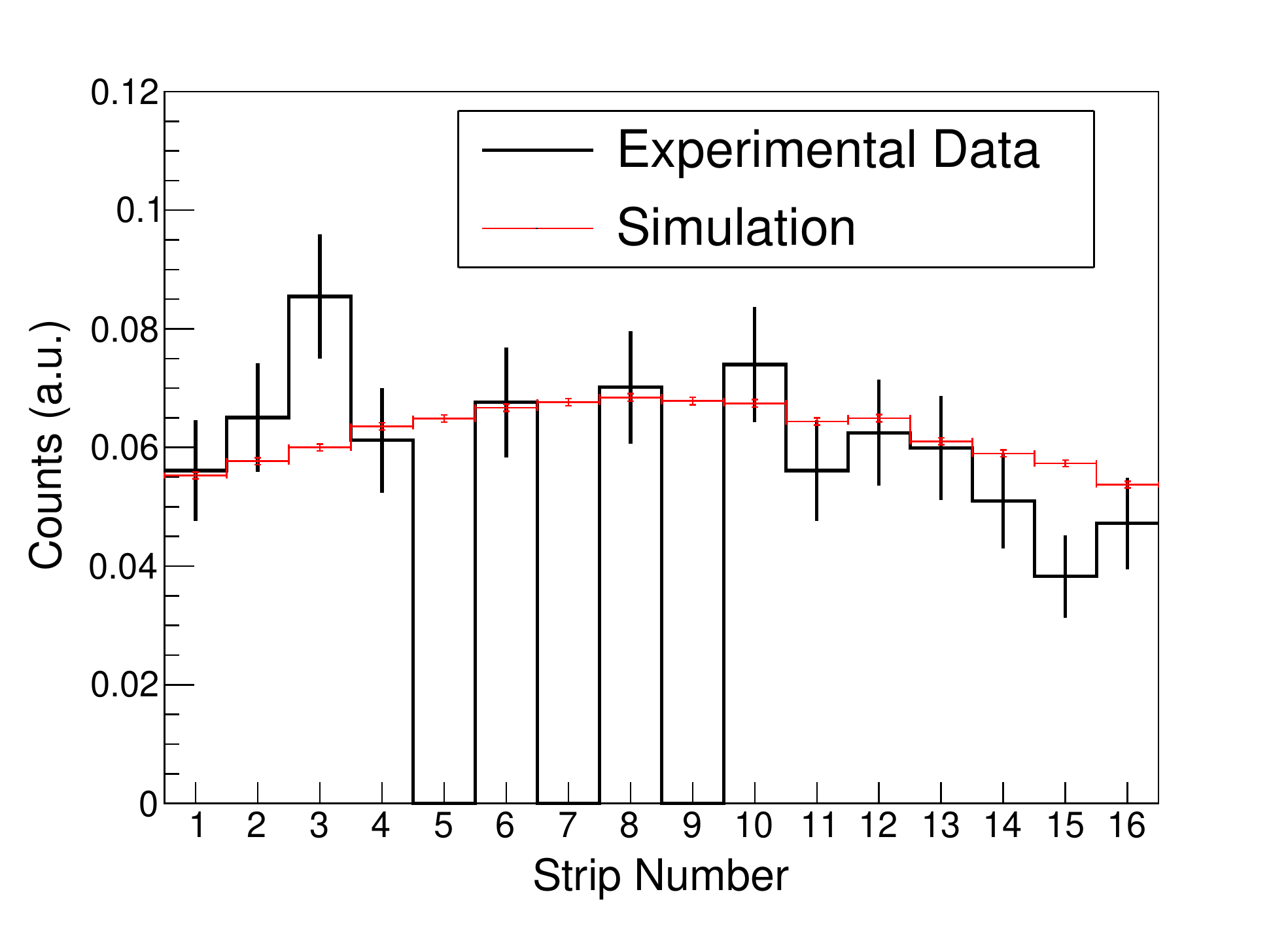}

\caption{Experimental counts at the horizontal strips of the DSSSD (black), compared with the simulated distribution of counts for protons emitted isotropically (red). Data are normalized so that the total number of counts (from all active strips) is one. Error bars of the experimental data correspond to statistical uncertainty.}

\label{fig:AngDistr1res}
\end{figure}

The angular distribution of the protons emitted from the 492.7 keV resonance was analyzed by means of the counts at individual strips in the DSSSD. Dedicated MCNP simulations were run using an isotropic proton distribution in the Center of Mass reference system. Protons were generated along the adenine sample with a spatial distribution given by the neutron beam profile at a distance of 19.75 m from the n\_TOF target and the proton energy corresponding to the center of the resonance. Figure \ref{fig:AngDistr1res} shows the count distribution in the strips of the detector in the horizontal direction that is more sensitive to the angular distribution. The experimental data is in agreement with a distribution arising from protons emitted isotropically. The observed result, compatible with isotropy, is in line with the spin adopted from compilations ($J=\frac{1}{2}$) for the first resonance. 

For the determination of the cross section (and resonance parameters) we also need to know the angular distribution from the second resonance at 644 keV. A count distribution compatible with isotropy was also found for the second resonance (though with fewer statistics), also in line with the $J=\frac{1}{2}$ from the Compilation in Ref. \cite{Mughabghab18}.

\subsection{R-Matrix Analysis}

To deduce the cross-section from the DSSSD yield, data were fitted using the R-matrix code\textsc{sammy} \cite{SAMMY}, applying the Reich-Moore approximation.

\begin{figure}[h!]

\centering
\includegraphics[width=.49\textwidth]{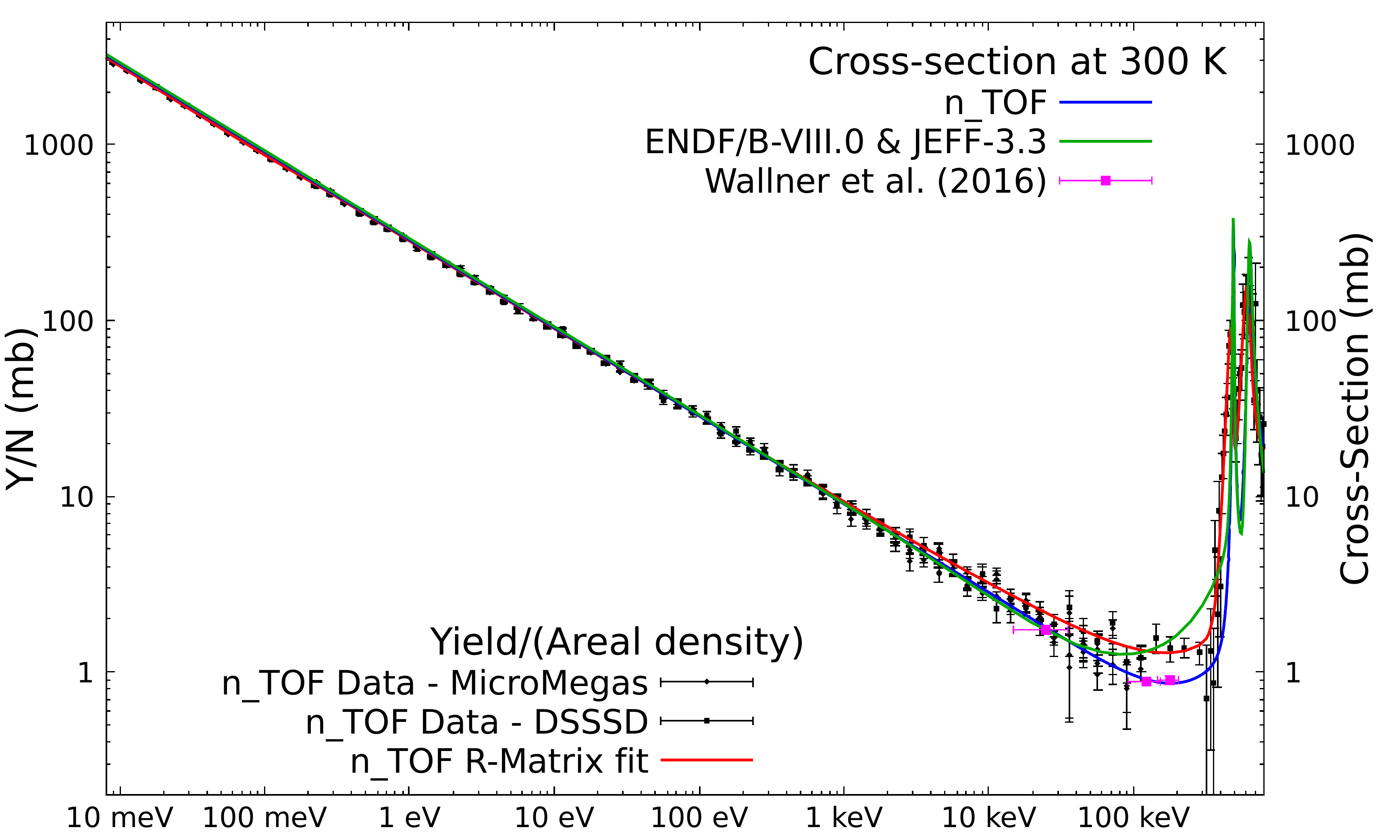}
\includegraphics[width=.49\textwidth]{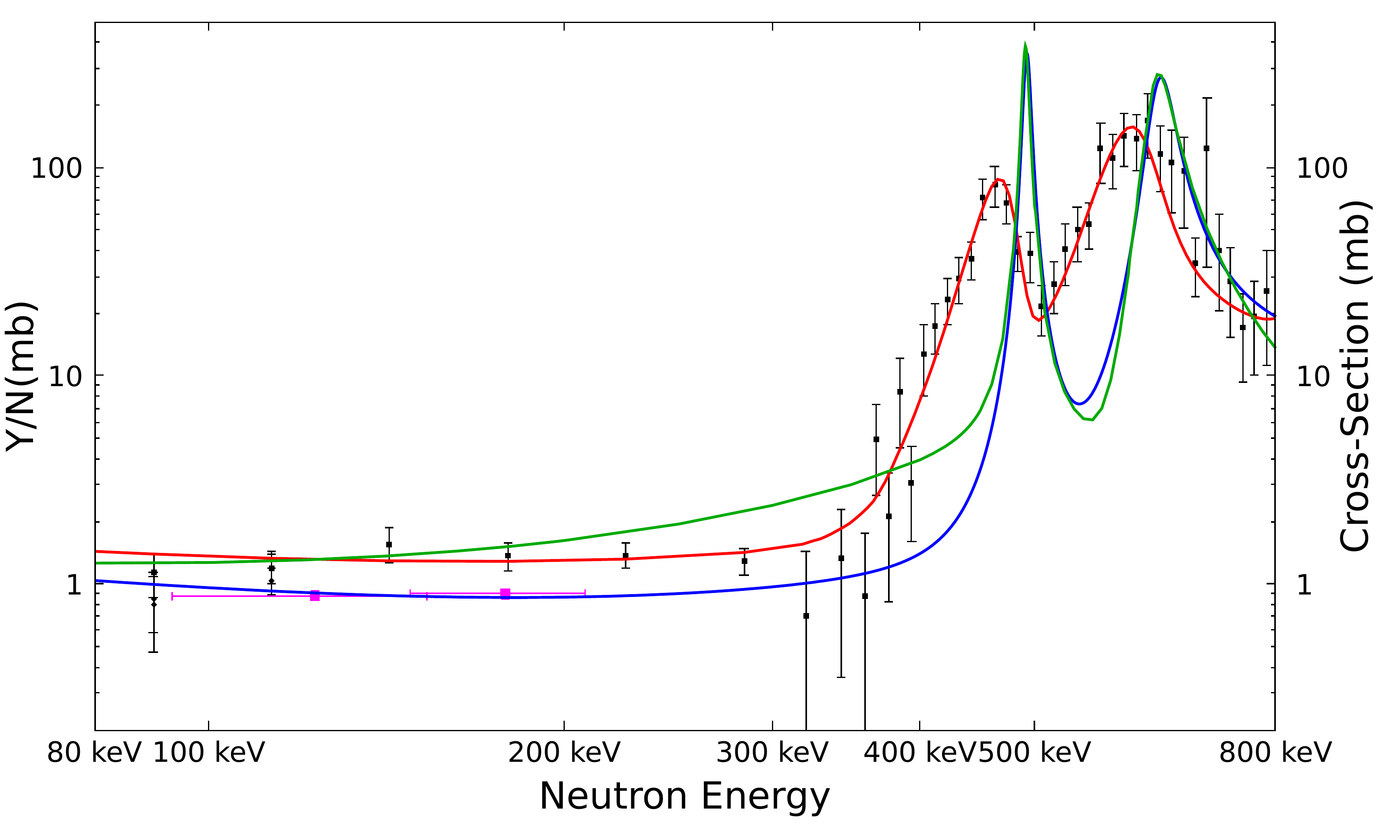}
\caption{The Yield data corrected for the areal density ($Y/N$) are shown in black. The red line is the \textsc{sammy} fit of the ($Y/N$) data, the blue line is the cross-section at 300 K deduced from parameters obtained with \textsc{sammy}. 
The integral measurements by Wallner \textit{et al.} at 25, 123 and 178 keV are shown in magenta. 
The ENDF/B-VIII.0 evaluation is included (in green). The upper panel shows the full range covered by this measurement, and the lower panel shows a detail of the resonance region.
}
\label{fig:Sammyfit}
\end{figure}

In order to illustrate the quality of the \textsc{sammy} fit to the yield data, Figure \ref{fig:Sammyfit} shows the experimental and fitted yield corrected for the sample areal density. In addition, the deduced cross-section (at 300 K) from the \textsc{sammy} is given. The ENDF/B-VIII.0 evaluation and the Wallner \textit{et al.} results are also shown for comparison.  Multiple scattering and self-shielding effects were found negligible given the small thickness of the adenine samples. A small constant background was considered in the fit and it was found to be 0.42 mb. It has no impact on derived resonance parameters and on the cross-section in the 1/v region. However, subtracting a constant background becomes relevant at neutron energies between about 50 and 300 keV if we want to reproduce simultaneously the data with the resonance parameters. Furthermore, subtracting this small background is only needed for the DSSSD data, since the MicroMegas yield is fully consistent with no background contribution. We would like to note that the actual background considered in DSSSD data must not necessarily be exactly constant over the whole fitted energies but only have a similar value to that required near 100 keV. As the deduced cross-section from n\_TOF in this region agrees perfectly with Wallner \textit{et al.} data, we have a high confidence that this region is treated correctly.  

In the present analysis, the $J^{\pi}$ of both resonances was assumed from Ref. \cite{Ajzenberg-Selove86}. The channel radii were taken as 5.5 fm; $\Gamma_\gamma$, $\Gamma_\alpha$ and $E_R$ were fixed to values from Ref. \cite{Mughabghab18}, while $\Gamma_n$ and $\Gamma_p$ were fitted.  The impact of the exact $\Gamma_\gamma$ and $\Gamma_\alpha$ on the fit is negligible provided that they are much smaller than the other widths. from Ref \cite{Mughabghab18}. The 432 keV resonance, reported in Ref. \cite{Mughabghab18} and present in the total cross-section of nitrogen \cite{Hinchey52}, was not observed neither in this measurement nor in any other papers reporting results at this energy from the \tssc{14}N(n,p) \cite{Johnson50} or the \tssc{14}C(p,n) reaction \cite{Gibbons59,Wang91}. Therefore, the $\Gamma_p$ of this resonance was fixed to zero. Bound states were included in the analysis in order to reproduce the 1/v behavior of this reaction below tens of keV. Their positions were again taken from Ref. \cite{Mughabghab18}, while $\Gamma_n$ and $\Gamma_p$ were adjusted. Other higher-lying resonances (up to 1.5 MeV) have also been included in the analysis, fixing all parameters to those in Ref. \cite{Mughabghab18}.

The experimental yield was consistently described in the range from 8 meV to 800 keV with the \textsc{sammy} fit. The resonance partial widths and resonance strengths, defined as $g\Gamma_n\Gamma_p/\Gamma$,  where $\Gamma= \Gamma_n + \Gamma_p + \Gamma_\alpha + \Gamma_\gamma$ is the total resonance width and the statistical factor $g=(2J+1)/[(2I+1)(2i+1)]$, with $J$, $I$ and $i$ the spins of the resonant state, the target ($1^+$) and the projectile ($\frac{1}{2}^+$), respectively, are listed in Table \ref{tab:tablerespar}.

\begin{table*}[h]
    \centering
        \begin{tabular}{lcccccc}
        \hline\hline
        $E_R$ (keV) & $J^\pi$ & $\Gamma_\gamma$ (eV) & $\Gamma_\alpha$ (keV)  & $\Gamma_n$ (keV) & $\Gamma_p$ (keV) & $g\Gamma_n\Gamma_p/\Gamma$ (keV)  \\ \hline
        432   & $\frac{7}{2}^+$  & $-$          &    $-$     & 1.86           & 0.0 & 0.0 \\
        492.7 & $\frac{1}{2}^-$ & $0.29$       &    $-$     & $1.90\pm0.15$  & $6.2\pm0.8$ & $0.48\pm0.04$    \\
        644   & $\frac{1}{2}^+$ & $4.2$  &  $0.0$   & $35\pm5$       & $9.2\pm0.7$ & $2.41\pm0.22$    \\ 
        837   & $\frac{1}{2}^+$ & $19.2$ &  $0.0$   &  $4.0$   & $400.9$ &          1.32       \\ 
        997   & $\frac{3}{2}^+$ &  $-$         &    $-$     &  $43.5$  & 0.8         &   0.52              \\
        1116  & $\frac{3}{2}^-$ &  $-$        &    $-$     &  $12.7$  &  4.4        &        2.18         \\
        1184  & $\frac{5}{2}^+$ &  $-$        &    $-$     &      1.3       &     $-$     &  $-$       \\
        1211  & $\frac{1}{2}^-$ &  $-$        &    0.26    & $14$       & $0.4$ &        0.13          \\
        1351  & $\frac{5}{2}^+$ &  $-$        &    0.4     & $19$       & $0.8$ &            0.75      \\
        1405  & $\frac{3}{2}^-$ &   $-$       &    1.8     &  $42$      & $11$    &            5.6      \\
        
        \hline\hline
\end{tabular}
    \caption{Parameters involved in the \textsc{sammy} fit of the first two \tssc{14}N(n,p)\tssc{14}C resonances. All values without error correspond to fixed parameters in the analysis: $J^\pi$ was taken from Ref \cite{Ajzenberg-Selove86}, $\Gamma_\gamma$, $\Gamma_\alpha$ and $E_R$ fixed using the expectation values from Ref. \cite{Mughabghab18}; $\Gamma_\alpha=0$  was used in the fitting of the 644 keV resonance, Ref. \cite{Mughabghab18} gives $\Gamma_\alpha<0.3 keV$. The 432 keV resonance was not observed and thus $\Gamma_p$ was set to zero. The (n,p) resonance strengths $g\Gamma_n\Gamma_p/\Gamma$ were derived from individual resonance parameters. Uncertainties correspond to the sum of statistical uncertainties (from \textsc{sammy} fit) and systematic uncertainties (from normalization of DSSSD data).}
    \label{tab:tablerespar}
\end{table*}

As mentioned above, the fit of the experimental data includes the effect of the EAR-2 Resolution Function, which is specially relevant in the resonance region, as clearly seen in the lower panel of Figure \ref{fig:Sammyfit}. The Resolution Function broadens the resonances but also shifts the maximum of the resonance in the yield towards lower energy. 

The n\_TOF cross-section describes well the 1/v behavior up to 50 keV. In contrast, the ENDF/B-VIII.0 evaluation shows a slight deviation from 1/v starting from $\approx$3 keV. In the resonance region, the n\_TOF cross-section is consistent within uncertainties with the ENDF/B-VIII.0 evaluation; in terms of the integrated cross-section of the resonance, between 450 and 550 keV, the n\_TOF reconstructed cross-section is 5.1 \% lower. 

A noteworthy situation appears at the low-energy tail of the first resonance (between about 50 and 450 keV), where our cross section differs significantly from the current ENDF/B-VIII.0 evaluation. However, the n\_TOF cross-section provides a much more consistent shape with the JENDL-5 evaluation in this range. Our shape in this region is also consistent with the data by Johnson et al. \cite{Johnson50}. Additionally, we are fully consistent with Wallner \textit{et al.} \cite{Wallner16} at 127 and 178 keV.  The disagreement between ENDF/B-VIII.0 and their cross-section led Wallner \textit{et al.} \cite{Wallner16} to propose a reduction on the strength of the first resonance by a factor of about 3.3. However, this proposal was based on the assumption that the cross-section energy dependence in ENDF/B-VIII.0 is correct in the region measured by Wallner \textit{et al.}. We thus conclude that the experimental data of Wallner \textit{et al.}, at 127 and 178 keV, as well as those of Johnson \textit{et al.} and Morgan at higher energies, are correct, in disagreement with the ENDF/B-VIII.0 evaluation in $E_n\approx3-450$ keV.


\subsection{Thermal Cross-section}

\begin{figure}[h!]

\centering
\includegraphics[width=.49\textwidth]{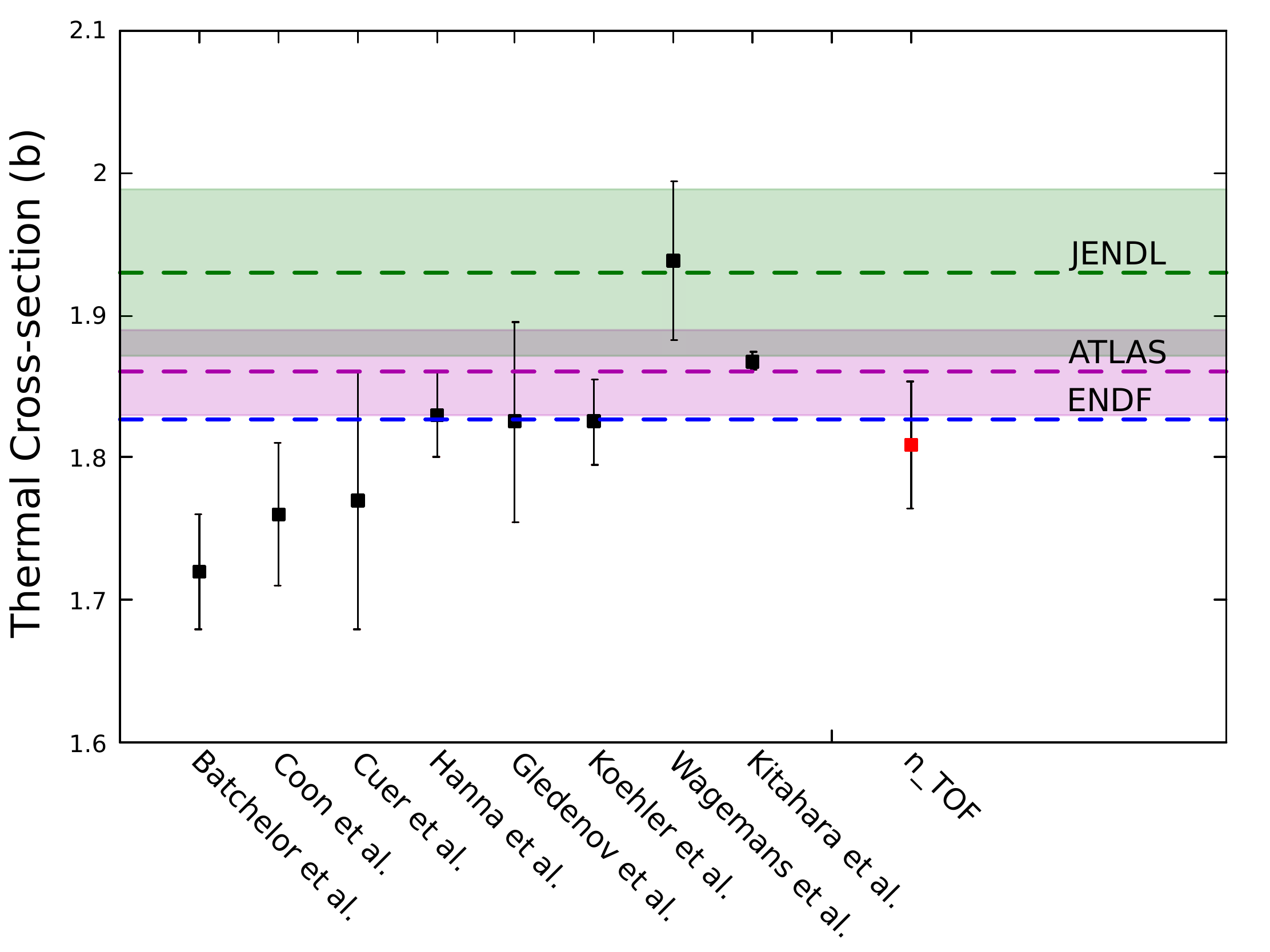}
\caption{ \tssc{14}N(n,p)\tssc{14}C thermal values. ENDF/B-VIII.0 and JEFF-3.3 (\textit{blue}), JENDL-5 (\textit{green}) and Atlas of Neutron Resonances (magenta, labelled ATLAS) \cite{Mughabghab18} are indicated with the dashed lines, with semi-transparent bands corresponding to uncertainty. Experimental values are marked with squares, our value is shown in red.
}
\label{fig:Thermal}
\end{figure}

The R-matrix fit of the corrected experimental yield ($Y/N$) allowed also the extraction of the thermal cross-section of the \tssc{14}N(n,p)\tssc{14}C reaction. It is found to be $1.809\pm0.045$ b. Figure \ref{fig:Thermal} compares this value with results of previous measurements and also with present evaluations. Our result is in agreement with ENDF/B-VIII.0 (JEFF-3.3), while it is lower than the JENDL-5 evaluation, which adopts the original value reported by Wagemans \textit{et al.} \cite{Wagemans00}. Our result is also compatible within uncertainties with the Compilation of the Atlas of Neutron Resonances by Mughabghab.
Compared to the most recent and very precise measurement by Kitahara \textit{et al.} \cite{Kitahara19}, our values is lower by about 1.3 standard deviations. We are also in good agreement with the measurements of Gledenov \textit{et al.} \cite{Gledenov93}, as well as with the thermal cross-section from the data by Koehler \textit{et al.} \cite{Koehler93}. It is lower, but only by about 1.2 standard deviations than results obtained by Wagemans \textit{et al.}, after renormalization of their value due to a change in the \tssc{235}U(n$_{th}$,f) cross section used as reference \cite{Carlson18}. The same procedure was followed for other previous measurements, for instance for Refs. \cite{Koehler93, Gledenov93}, measured relative to \tssc{6}Li(n,t). 
    
\subsection{Maxwellian Averaged Cross-Section}

Maxwellian Averaged Cross-Sections (MACS) have been calculated for thermal energies in the range of $kT$ = 5-100 keV, using the cross-section obtained from \textsc{sammy}. The uncertainties are estimated considering a 2.5 \% uncertainty in the whole 1/v region, and the uncertainties in the resonance parameters as given in Table \ref{tab:tablerespar}. The results are listed in Table \ref{tab:tablemacs}, together with values given by Wallner \textit{et al.} \cite{Wallner16} and those derived from the ENDF/B-VIII.0 evaluation cross-section. All three MACS agree within uncertainties for $kT$ below about 15 keV. 
For higher $kT$ the different behavior of the tail of the first resonance results in differences in MACS from different sources. Specifically, the higher cross-section considered in ENDF/B-VIII.0 above a neutron energy of about 40 keV makes our MACS smaller. The Wallner \textit{et al.} MACS values above $kT\approx$ 60 keV are strongly impacted by their suggestion related to the strength of the first resonance.




\begin{table*}[h]
    \centering
        \begin{tabular}{lcccccc}
        \hline\hline
        kT (keV) & Wallner \textit{et al.} & ENDF/B-VIII.0  & n\_TOF \\ \hline
        5 &   $3.78\pm0.06$ & 3.81 & $3.91\pm0.10$\\
        8 &   $3.12\pm0.05$ & 3.01 & $3.09\pm0.08$\\
        10 &  $2.89\pm0.05$ & 2.70 & $2.76\pm0.07$\\
        15 &  $2.47\pm0.04$ & 2.26 & $2.26\pm0.06$\\
        20 &  $2.21\pm0.04$ & 2.02 & $1.97\pm0.05$\\
        23 &  $2.09\pm0.04$ & 1.93 & $1.84\pm0.04$\\
        25 &  $2.03\pm0.04$ & 1.88 & $1.77\pm0.04$\\
        30 &  $1.93\pm0.04$ & 1.80 & $1.63\pm0.04$\\
        40 &  $1.85\pm0.05$ & 1.75 & $1.47\pm0.03$\\
        50 &  $1.83\pm0.06$ & 1.86 & $1.46\pm0.03$\\
        60 &  $1.84\pm0.07$ & 2.23 & $1.69\pm0.04$\\
        80 &  $1.84\pm0.08$ & 3.96 & $3.14\pm0.10$\\
        100 & $1.83\pm0.08$ & 6.92 & $5.83\pm0.20$\\  \hline\hline
\end{tabular}
    \caption{Maxwellian Averaged Cross Sections derived from the n\_TOF cross-section, compared with the previous calculation by Wallner \textit{et al.} \cite{Wallner16} and the value based on theENDF/B-VIII.0 evaluation cross-section \cite{ChadwickEval99}. Total uncertainties are given. }
    \label{tab:tablemacs}
\end{table*}

\section{SUMMARY AND CONCLUSIONS}

A new measurement of the \tssc{14}N(n,p)\tssc{14}C reaction has been performed at the EAR-2 of the n\_TOF facility at CERN. The measurement provides nuclear data from sub-thermal to the resonance region for the first time, spanning from 8 meV to 800 keV. The cross-section is obtained via fitting the experimental yield using the \textsc{sammy} code.  The obtained thermal cross-section is 1.809$\pm$0.045 b, in good agreement with ENDF/B-VIII.0 and JEFF-3.3, lower than the JENDL-5 evaluation and by slightly more than one standard deviation lower than values reported from the two most recent dedicated measurements by Wagemans \textit{et al.} \cite{Wagemans00} and Kitahara \textit{et al.} \cite{Kitahara19}. Our dependence of the cross-section on neutron energy then starts to differ from ENDF/B-VIII.0 and JEFF-3.3 evaluations above about 3 keV. The 1/v cross-section dependence is observed for all energies up to about 50 keV and we observe significant disagreement with respect to these evaluations between about 50 and 450 keV, at the low-energy tail of the first resonance at 492.7 keV. On the other hand, our cross-section in this range nicely reproduces the cross sections at 25, 123 and 178 keV obtained by Wallner \textit{et al.} \cite{Wallner16}. Our resonance integrals of the first two resonances at 492.7 and 644 keV are consistent with the values reported by Morgan and adopted by evaluations. These data could promote new evaluations of the cross-section of the \tssc{14}N(n,p) reaction.  With our cross-section data, a calculation of the MACS is carried out and the results show a good agreement with MACS deduced from evaluations at $kT$ below 15 keV with discrepancies at higher $kT$.



\begin{acknowledgments}
We thank Mr. Wilhelmus Vollenberg for the preparation of the Adenine samples.

For the purpose of open access, the author has applied a Creative Commons Attribution (CC BY) licence to any Author Accepted Manuscript version arising from this submission.

This work was partially supported by Spanish Ministerio de Ciencia e Innovación (PID2020-117969RB-I00), Junta de Andalucía (FEDER Andalucia 2014–2020) projects P20-00665 and B-FQM-156-UGR20. This work was also supported by the UK Science and Facilities Council (ST/M006085/1, ST/P004008/1 ), and the European Research Council ERC-2015-StG Nr. 677497. Also by the funding agencies of the n\_TOF participating institutes.

P. T. acknowledges support from the Spanish Ministry of Science, Innovation and Universities under the FPU grant FPU17/02305.
\end{acknowledgments}

\bibliography{refs}

\end{document}